\documentclass[12pt]{iopart}
\usepackage{iopams}

\usepackage{graphicx}

\newtheorem{theor}{Theorem}
\newtheorem{prop}{Proposition}

\makeatletter
\def\sgn{\mathop{\operator@font sgn}\nolimits}
\makeatother

\begin{document}

\title[Ground states of spin polygons]{Ground states of classical spin polygons:
Rigorous results and examples}

\author{ W Florek$^1$, H-J Schmidt$^2$, and K Ja\'sniewicz-Pacer$^1$}

\address{$^1$ Faculty of Physics and Astronomy, Adam Mickiewicz University,
61-614 Pozna\'n, Poland}

\address{$^2$ Fachbereich Mathematik/Informatik/Physik, Universit\"{a}t
Osnabr\"{u}ck, D-49069 Osnabr\"{u}ck, Germany}

\ead{hschmidt@uos.de}

\begin{abstract}
We present a~comprehensive and rigorous analysis of the lowest energy
configurations (LECs) of classical spin polygons characterized by arbitrary
couplings between neighboring spin sites. Our study shows that these ground
states exhibit either collinear or coplanar arrangements, which allows us to
determine the precise boundaries between these two phases. By simultaneously
applying a~spin flip and a~bond inversion, we simplify the LEC problem and
reduce it to a~specific scenario with predominantly ferromagnetic (FM) bonds
and a~single antiferromagnetic (AFM) bond. Hence, competing interactions are
always present, but, nevertheless, in the well-defined ranges of the system
parameters the collinear LEC is realized. The difference angles between
neighboring spins within the LEC can be captured by a~single Lagrange
parameter. We analytically investigate its dependence on the AFM bond and arrive
at revealing results. Similarly, we can analyze the energy of the LEC, which
shows a~pronounced maximum as a~function of AFM bond. To illustrate our
findings, we give various examples that clearly demonstrate these results.
\end{abstract}

%
\vspace{2pc}
\noindent{\it Keywords}: classical spin systems, competing interactions,
phase diagram
%

%
%
%

\section{Introduction}\label{Intro}
For years single molecule magnets have occupied a~prominent place in chemical
and physical research, both experimental and theoretical. Recently, increasingly
larger molecules containing, in addition to transition metals, also lanthanide
ions have been synthesized \cite{Bani18,Maje18,Crai19,Fu2020,Antk21,Shuk21,
Tzio23,Jin23}. In many cases, dimensions of the corresponding eigenproblem for
a~\emph{quantum\/} spin system in question are definitely too high to apply any
of existing exact diagonalization method, even when employing all symmetries.
Given this situation, a~~\emph{classical\/} treatment yields a~first approximation for
individual spins numbers $s\gg1$, which is often astonishingly good (cf.\
\cite{Dmit19}). Therefore, results obtained in the classical limit may yield,
for example, approximation schemes for quantum problems or impose some
restrictions on their prospective solutions \cite{Enge06,Kons18,Schu21}. On the
other hand, classical infinite lattices and finite spin systems are also worth
investigating on their own and without direct relationships to their quantum
analogues \cite{Ueda17,Gome18,Anis19,GS23}.

This paper is devoted to classical counterparts of quantum ringshaped homo-
and heterometallic spin systems, which play an important role in the realm of
single molecule magnets. They are synthesized using various magnetic centres,
differ in nuclearity and multifarious magnetic interactions are observed. This
profusion of samples not only results in abundant amount of experimental data
but also induces many theoretical works and possible applications (e.g.\
molecular qubits or low-temperature cooling) \cite{Furr13,Timc13,Shar14,McIn15,
Ghir15,Bake16,Garl16,Antk19,Schn19,Maty21,Poin23,Chie24}. The relatively simple
system of interactions (each spin is coupled to its two nearest neighbours only)
allows to obtain very general interesting and important results.

Since there is only one cycle the concept of frustration function proposed by
Toulouse \cite{Toul77} for spin glasses can be extrapolated to our case: A~polygon
is considered as frustrated if the number of antiferromagnetic (AFM) exchange integrals is odd.
This is equivalent to another criterion based on the concept of competing interactions.
They are present when there is a~competition between different
interactions and not all of bonds can be satisfied simultaneously \cite{Diep04}.
In the other words, the minimum of the system energy is greater than a~sum of
the minimal one-bond energies. For a~system of classical spins without competing
interactions the lowest energy configuration (LEC) has to be collinear. Hence,
the coplanar or spatial LECs indicate the presence of competing interactions.
However, these statements do not exclude that the collinear LECs are realized
in the presence of competing interactions (see, for example, \cite{Shas81,
Voig95,Rich96,Kami15,Flor16,Flor19,Schm22}). As a~rule this feature is observed
in the well-determined range of exchange integrals.

The aim of this work is to analyze the LECs for classical spin polygons as
closely and as fully as possible. The paper is organized as follows. In
\sref{Poly} we define the class of systems we will study, the classical $N$-spin
polygons. The first Theorem states that the ground states of these systems are
either collinear or coplanar. Furthermore, we introduce the $N$~difference
angles~$\psi_\mu$ between adjacent spin vectors as coordinates to describe the
ground states, subject to the constraint that their sum must be an integer
multiple of $2\pi$. In the next \sref{GenProp} we use a~well-known tool, the
simultaneous spin flip and bond inversion, to reduce the LEC problem to the
special case where all bonds are ferromagnetic (FM) except for a~single
antiferromagnetic (AFM) bond~$\alpha_0$. Furthermore, it is shown that all
difference angles can essentially be written as functions of a~Lagrange
parameter~$\beta$, due to the constraint mentioned above. It is also shown that
only two collinear states, symbolized by $\uparrow\ldots\uparrow$ and
$\uparrow\ldots\downarrow$, are possible ground states of polygons.

Next, we address the obvious problem for which values of the AFM bond~$\alpha_0$
the LEC will be collinear. It turns out that our approach requires a~restriction
to the interval $0<\alpha_0<|\alpha_\tau|$, where~$\alpha_\tau$ denotes the
weakest FM bond. For this interval, called `regular domain', we prove in
\sref{RegDom} the existence of a~critical value~$\alpha_{\rm (c)}$ so that
$0<\alpha_0\le\alpha_{\rm (c)}$ corresponds to the collinear phase and
$\alpha_{\rm (c)}<\alpha_0\le|\alpha_\tau|$ corresponds to the coplanar phase.
The critical value~$\alpha_{\rm (c)}$ can be expressed by the harmonic mean of
absolute values of the FM bonds. In the regular domain, the difference angles
and the Lagrange parameter~$\beta$ depend analytically on~$\alpha_0$, while for
$\alpha_0=\alpha_{\rm (c)}$ a~square root expansion of these quantities in terms
of $\alpha_0-\alpha_{\rm (c)}$ applies. The energy has a~single maximum in the
regular domain.

The results for the regular domain can be transferred to the `complementary
domain', which is given by $|\alpha_\tau|<\alpha_0<\infty$, using a~spin-flip
transformation. This is done in \sref{ComDom}. We also obtain striking values
in the complementary domain
for $\alpha_0=\alpha_{\rm (bm)}$, where a~maximum of~$\beta(\alpha_0)$ can be
assumed and for $\alpha_0=\alpha_{\rm (c')}$, where the second phase transition
to a~collinear phase can take place. However, whether this happens depends
on~$\alpha_\tau$, and this dependence is further investigated in \sref{PD}.
There, a~`phase diagram' in the $\alpha_0$--$|\alpha_\tau|$ plane is developed
and discussed. Examples illustrating these results are given in \sref{Ex}. The
paper concludes with a~summary in \sref{Sum}.

The proofs for most of the Propositions and Theorems as well as the discussion
of the limiting case between the regular and complementary domains are included
in the appendices to help readers understand the results in the main text without
going into mathematical details.

\section{Classical spin polygons}\label{Poly}
The systems considered in this work will be referred to as ``polygons".
In graph-theoretical terms \cite{GraphTh} the system is assumed to be a
cycle graph, i.~e., a connected graph  such that all vertices have exactly two neighbours.
It is thus possible to label the $N>2$ vertices by the numbers $\mu=0,1, \ldots, N-1$
such that the two neighbours of the vertex $\mu$ have the numbers $\mu-1$ and $\mu+1$,
understood $\bmod\, N$. This labelling is only unique up to cyclic permutations
and the inversion $(0,1,\ldots,N-1)\mapsto (N-1,N-2,\ldots,0)$.
Later, we will make use of this freedom to choose the labelling.

Moreover, the vertices are carriers of classical
spins~$\bi{s}_\mu\in \mathbb{R}^3$ with $|\bi{s}_\mu|=1$ for all $0\le\mu<N$. With
the assumed labelling scheme all expressions with these indices are
understood $\bmod\, N$, unless otherwise stated. The vectors~$\bi{s}_\mu$ can be interpreted
as rows of the $N\times 3$-dimensional matrix~$\mathbf{S}$, where an
element~$S_{\mu\,j}$, $j=1,2,3$, is the $j$th component of a~vector~$\bi{s}_\mu$.

There are $N$~edges (bonds) $e_\mu:= (\mu,\mu+1)$ with weights (exchange
integrals) $\alpha_\mu\neq 0$, such that the energy can be written as
\begin{equation}\label{EnPoly}
  E(\mathbf{S})=\sum_{\mu=0}^{N-1} \alpha_\mu \,\bi{s}_\mu\cdot\bi{s}_{\mu+1}
    =\sum_{\mu=0}^{N-1}E_\mu,
\end{equation}
where $E_\mu$ denotes the energy of a~bond~$e_\mu$. The LEC is a~spin
configuration globally minimizing the energy~$E(\mathbf{S})$. Let for chosen $0\le\mu<N$
all vectors~$\bi{s}_\nu$, $0\le \nu<N$ and $\nu\ne\mu$, kept fixed then this
particular spin~$\bi{s}_\mu$ has to minimize the one-spin energy
\begin{equation}\label{OSEPoly}
  \mathcal{E}_\mu(\mathbf{S})
    =\bi{s}_\mu\cdot(\alpha_{\mu-1}\bi{s}_{\mu-1}+\alpha_\mu\bi{s}_{\mu+1})
    =E_{\mu-1}+E_\mu.
\end{equation}
Hence this vector has to satisfy the so-called stationary state equation
\cite{Schm03,Schm17a}
\begin{equation}\label{ssep}
  -\kappa_\mu\bi{s}_\mu=\alpha_{\mu-1}\bi{s}_{\mu-1}+\alpha_\mu\bi{s}_{\mu+1}.
\end{equation}
Alternatively, the $\kappa_\mu$ can be regarded as Lagrange parameters, which
result from the minimization of the energy under the constraints
$|\bi{s}_\mu|=1$. Note the relation
$\mathcal{E}_\mu(\mathbf{S})=-\kappa_\mu$ which follows by
taking the scalar product of (\ref{ssep}) with $\bi{s}_\mu$.
It is assumed that all couplings
$\alpha_\mu\ne 0$, since otherwise the spin polygon degenerates into one or more
spin chains which have trivial collinear LECs. This is the only restriction, so in
the general case this system has no spatial symmetry.

The simple structure of magnetic interactions in the model considered allows to
prove the following basic theorem.
\begin{theor}\label{MTh}
Assume a~classical spin system of $N$~spin vectors~$\bi{s}_\mu$, with
$|\bi{s}_\mu|=1$ for all $0\le\mu<N$, such that each spin~$\bi{s}_\mu$ is only
coupled to spins $\bi{s}_{\mu\pm1}$ with non-zero exchange integrals
$\alpha_\mu$ and $\alpha_{\mu-1}$, respectively, and the periodic boundary
conditions $N+\mu\equiv\mu$ are adopted. Then the lowest energy configuration of
this system is either collinear or coplanar.
\end{theor}

\noindent\textbf{Proof.}~ The stationary state equation~(\ref{ssep}) can be
rewritten as
\[
  \alpha_{\mu-2}\bi{s}_{\mu-2}+\kappa_{\mu-1}\bi{s}_{\mu-1}
    +\alpha_{\mu-1}\bi{s}_\mu=\mathbf{0},
\]
so any three consecutive spins constitute a~set of linearly dependent non-zero
vectors, therefore they are either collinear or coplanar. It follows from this
formula that
\begin{equation}\label{InProof1}
  \bi{s}_\mu=
    -(\alpha_{\mu-2}\bi{s}_{\mu-2}+\kappa_{\mu-1}\bi{s}_{\mu-1})/\alpha_{\mu-1}.
\end{equation}

\noindent \textsl{Case~(i)}: $\bi{s}_0$ and $\bi{s}_1$ are collinear.~ In this
case $\bi{s}_1=\pm\bi{s}_0$. Hence,
$\bi{s}_2=-(\alpha_0\pm\kappa_1)\bi{s}_0/\alpha_1$ is also collinear with
$\bi{s}_0$. If for any $2< \mu<N$ spins $\bi{s}_{\mu-2}$ and $\bi{s}_{\mu-1}$
are collinear with $\bi{s}_0$, then, by~\eref{InProof1}, also $\bi{s}_\mu$ is
collinear with~$\bi{s}_0$. Therefore, by induction over~$\mu$, all spins
$\bi{s}_\mu$, $0<\mu<N$, are collinear with $\bi{s}_0$, thus the lowest energy
configuration is collinear.
\medskip

\noindent \textsl{Case~(ii)}: $\bi{s}_0$ and $\bi{s}_1$ are coplanar.~ Let~$P$
denote the two-dimensional subspace of $\mathbb{R}^3$ spanned by
$(\bi{s}_0,\bi{s}_1)$. The formula~\eref{InProof1} ensures that $\bi{s}_2\in P$.
Generally, if $\bi{s}_{\mu-2}$ and $\bi{s}_{\mu-1}\in P$, then also
$\bi{s}_\mu\in P$. Therefore, by induction over~$\mu$, all spins
$\bi{s}_\mu\in P$, $1<\mu<N$, thus the lowest energy configuration is coplanar.
\hfill$\square$
\bigskip

If considerations are restricted to the LECs then, according to this Theorem,
all vectors $\bi{s}_\mu$ can be assumed to lie in a two-dimensional subspace
$P\subset \mathbb{R}^3$ and written in the form $\bi{s}_\mu=(\cos\phi_\mu,\sin\phi_\mu)$,
where the angles~$-\pi<\phi_\mu\le \pi$ are measured with respect to an arbitrarily
chosen direction ${\mathbf e}\in P$. The degeneracy of the LEC with respect
to planar rotations can be removed by another representation characterized by a~sequence
$\bPsi_0=(\psi_0,\psi_1,\ldots,\psi_{N-1})$ of difference angles between successive spins
with $\psi_\mu:=\phi_{\mu+1}-\phi_\mu$.
This representation will be independent of ${\mathbf e}$.
It follows that the difference angles lie in the larger interval
\begin{equation}\label{psirange}
 -2\pi < \psi_\mu \le 2\pi ,\qquad {\rm for\; all}\quad 0\le \mu < N
 \;,
\end{equation}
and that their ``telescope sum" vanishes
\begin{equation}\label{Sumpsi}
 \sum_{\mu=0}^{N-1}\psi_\mu =0
 \;.
\end{equation}
As in the proof of theorem \ref{MTh} we can argue that
$\psi_\mu=\pm n \pi$  for $n=0,1,2$  for some $0\le \mu <N$
implies that the LEC is collinear. Hence the following holds
\begin{equation}\label{psi0coplanar}
  \psi_\mu\neq\pm n \pi \quad {\rm for}\; n=0,1,2\; {\rm and\;all }\; 0\le \mu<N
  \; {\rm if\; the\; LEC\; is\; coplanar}.
\end{equation}
By adding or subtracting $2\pi$, if necessary, we may re-define the $\psi_\mu$
such that they lie in the smaller interval
\begin{equation}\label{Newpsirange}
 -\pi < \psi_\mu \le \pi \qquad {\rm for\; all}\quad 0\le \mu < N
\;.
\end{equation}
However, this has the side effect that the sum of the re-defined $\psi_\mu$
no longer vanishes but satisfies
\begin{equation}\label{AngCons}
  \sum_{\mu=0}^{N-1}\psi_\mu=2k\pi,\qquad k\in\mathbb{Z}
  \;.
\end{equation}
Analogously to \eref{psi0coplanar} we may state that
\begin{equation}\label{newpsi0coplanar}
  \psi_\mu\neq 0,\pi \; {\rm for\;all }\; 0\le \mu<N  \; {\rm if\; the \;LEC \;is \;coplanar}.
\end{equation}
After eliminating $\psi_0$ by means of
\begin{equation}\label{Psi0}
  \psi_0=2k\pi-\sum_{\mu=1}^{N-1}\psi_\mu=:2k\pi-\Psi
\end{equation}
the energy of a~system in question can be written as a~function of a~sequence
$\bPsi=(\psi_1,\psi_2,\ldots\,\psi_{N-1})$ in the following way
\begin{equation}\label{EnbyPsi}
  E(\mathbf{S})\equiv E(\bPsi)
    =\alpha_0\cos\Psi+\sum_{\mu=1}^{N-1}\alpha_\mu\cos\psi_\mu.
\end{equation}
Since $\cos$ is an even function it is evident that after a~\textit{global\/} transformation
$\psi_\mu\mapsto-\psi_\mu$ for $1\le\mu<N$, so also $\Psi\mapsto-\Psi$,
the energy does not change, i.~e., $E(\bPsi)=E(-{\bPsi})$. This
transformation can be achieved by an inversion of the labelling of the spins and
is different from the other global transformation
$\bi{s}_\mu\mapsto-\bi{s}_\mu$, $0\le\mu<N$, under which the angles~$\psi_\mu$
remain constant $\bmod \,2\pi$.  We will use the freedom of
choosing the labelling of the spins to fix the sign of~$\sin \Psi$ as non-negative:
\begin{equation}\label{Signsinpsi}
\sin \Psi\ge 0
  \;.
\end{equation}
This choice will greatly simplify the following considerations.

\section{General properties of the LEC}\label{GenProp}

\begin{figure}[tb]
\begin{center}
  \includegraphics[width=12cm]{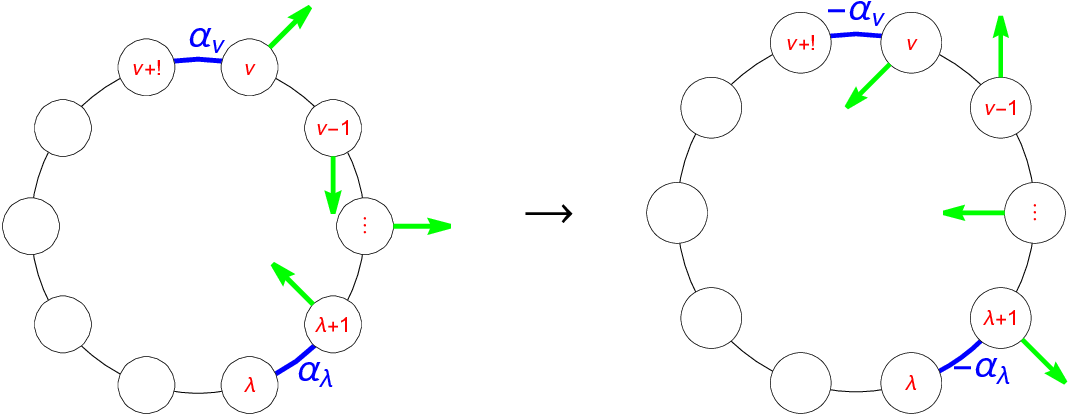}
\end{center}
\caption{
Illustration of a spin flip operation $\bi{s}_{\mu}\mapsto -\bi{s}_{\mu},\; \mu=\lambda+1, \ldots, \nu$
accompanied by a sign reversion of the pair of bonds $\alpha_\mu \mapsto -\alpha_\mu,\; \mu=\lambda, \nu$
such that $\lambda \neq \nu$.
By this operation the energy \eref{EnPoly} is left unchanged.
}
\label{figrr}
\end{figure}

By flipping a~sequence of spins
$\bi{s}_\mu$, $\mu=\lambda+1,\lambda+2,\ldots,\nu-1,\nu$,
and reversing the signs of the exchange integrals $\alpha_\lambda$ and
$\alpha_{\nu}$ the energy of a~polygon is left unchanged
(cf.\ \cite{Toul77,Hara53,ThMag,Giam11}, see Figure \ref{figrr}.
By repeating this operation every even
number of antiferromagnetic bonds  can be transformed into ferromagnetic ones.
Thus there are two cases depending on whether the original polygon has
an even or an odd number of antiferromagnetic bonds $N_{\rm AFM}$.

In the first case a~polygon without competing interactions and hence
an even $N_{\rm AFM}$  can be transformed into a~system
in which all couplings are ferromagnetic (FM), i.e., $\alpha_\mu<0$ for all
$0\le\mu<N$. Such system has the LEC energy
\begin{equation}\label{PolFM}
  E_{\rm LEC}^{\rm FM}=\sum_{\mu=0}^{N-1} \alpha_\mu
    =-\sum_{\mu=0}^{N-1} |\alpha_\mu|
\end{equation}
and all bonds are satisfied with $E_\mu=-|\alpha_\mu|$.
The ground state of the FM polygon is the fully aligned
state $\uparrow\ldots \uparrow$. By reversing the spin flip
transformation we obtain a collinear ground state with the
structure $\uparrow\downarrow$ at the AF bonds.
Thus the LEC of these polygons is well understood and we can
exclude them from our subsequent considerations.

Hence, in the following parts of the present work we will only
consider the second class of polygons with competing interactions
and hence an odd $N_{\rm AFM}$ such that
\begin{equation}\label{PolCI}
   E_{\rm LEC} > -\sum_{\mu=0}^{N-1} |\alpha_\mu|
   \;.
\end{equation}
This means that at least one bond is not satisfied, i.~e., $E_\mu>-|\alpha_\mu|$ for at
least one $0\le\mu<N$. In this case, by the above mentioned procedure
of flipping spins and reversing signs of the
appropriate bonds the polygon can be transformed to a~cycle
with only one AFM bond. Using the freedom of performing
a cyclic permutation of the indices when labelling the spins,
this unique AFM bond is chosen as~$\alpha_0>0$, such that the other bonds are
ferromagnetic, i.~e.,
\begin{equation}\label{FMAssum}
  \alpha_\mu <0,\qquad {\rm for\;\; all}\quad 0<\mu<N.
\end{equation}
It is also assumed that, except for \sref{PD}, these values are fixed,
whereas $\alpha_0$ is varying and all quantities discussed are considered as
functions of this variable.

Obviously, $\bPsi\mapsto E(\bPsi)$ is a~smooth function from the
$N-1$-dimensional torus $T^{N-1}$ to $\mathbb{R}$.
Since $T^{N-1}$ is compact, $E$ assumes its \emph{absolute\/}
minimum at some point $\bPsi=\widehat{\bPsi}$. Then the following necessary conditions for
a~\emph{local\/} minimum are satisfied:
\begin{equation}\label{MinCond1}
  \left.\frac{\partial E(\bPsi)}{\partial\psi_\mu}
    \right|_{\bPsi=\widehat{\bPsi}}=0, \qquad \mbox{for all}\quad 0<\mu<N.
\end{equation}
Applying this to \eref{EnbyPsi} the following equalities are obtained
\begin{equation}\label{MinCond1eval}
  \frac{\partial E(\bPsi)}{\partial\psi_\mu}
    =-(\alpha_0\,\sin\Psi+\alpha_\mu\,\sin\psi_\mu)
    = -\alpha_0\,\sin\Psi+|\alpha_\mu|\,\sin\psi_\mu=0,
\end{equation}
for all $0<\mu<N$. Hence we have proven the following:

\begin{prop}\label{propbeta}
For every LEC the quantity
\begin{equation}\label{defBeta}
  \beta:= \alpha_0\,\sin\Psi = \left|\alpha_\mu\right|\,\sin\psi_\mu
\end{equation}
will be independent of $\mu$ for all $0<\mu<N$.
\end{prop}
Note that, according to this Proposition, $\alpha_0>0$ and $\sin\Psi\ge 0$
by \eref{Signsinpsi} imply $\beta\ge 0$
and, further, $\sin\psi_\mu\ge 0$ for all $0< \mu <N$.
Hence, by virtue of \eref{Newpsirange}, we conclude
\begin{equation}\label{psimuge0}
 0\le \psi_\mu \le \pi \qquad {\rm for\;\; all}\quad 0<\mu<N.
\end{equation}

Alternatively, $\beta$~can be understood as the Lagrange parameter of the
minimization problem for~$E(\bPsi_0)$ subject to the constraint~\eref{AngCons}.
Hence we will refer to~$\beta$ simply as `the Lagrange parameter' in what
follows.

In order to get an intuitive idea of $\beta$~we may consider the
following mechanical model of the spin polygon. The $\bi{s}_\mu$ are viewed as
the positions of $N$ particles constrained to a unit circle and \eref{EnPoly}
is the potential energy of this mechanical system. Hence one can define the force
\begin{equation}\label{fmu}
    \bi{f}_\mu= - \nabla_{\bi{s}_\mu} E
    = -\alpha_{\mu-1}\bi{s}_{\mu-1}-\alpha_\mu\bi{s}_{\mu+1}
    \;,
\end{equation}
upon the $\mu$-th particle and the corresponding torque
\begin{equation}\label{nmu}
    \bi{n}_\mu= \bi{s}_\mu\times \bi{f}_\mu =
    -\alpha_{\mu-1}\bi{s}_\mu\times\bi{s}_{\mu-1}
    -\alpha_\mu\bi{s}_\mu \times\bi{s}_{\mu+1}
    \;.
\end{equation}
For an LEC this torque $\bi{n}_\mu$ must vanish for all $0\le \mu <N$
and hence the absolute value of
$-\alpha_\mu\bi{s}_\mu \times\bi{s}_{\mu+1}$ must be constant throughout the
spin polygon. The latter can be written in the form
\begin{equation}\label{ansmu}
\left|-\alpha_\mu\bi{s}_\mu \times\bi{s}_{\mu+1}\right|
= \left| \alpha_\mu \sin \psi_\mu\right| = |\beta|
\;.
\end{equation}
Therefore, according to this mechanical model, the Lagrange parameter $\beta\ge 0$
can be understood as the absolute value of the constant torque which the $\mu+1$-th
particle exerts on the $\mu$-th one in the LEC.

The total domain $0<\alpha_0<\infty$ is divided into two intervals,
$(0,|\alpha_\tau|)$ and $(|\alpha_\tau|,\infty)$, where the threshold is defined
as
\begin{equation}\label{AlThr}
  |\alpha_\tau|:=\min_{0<\mu<N} |\alpha_\mu|.
\end{equation}
The limit point $\alpha_0=|\alpha_\tau|$ has to be treated separately,
see \ref{Limit}. As the notation suggests, this minimum will be attained by one
or several coupling constants of which we have selected an arbitrary one
$|\alpha_\tau|$ with $0<\tau<N$.

The first interval is referred to below as the \emph{regular domain\/} and the
second one as the \emph{complementary domain}. This subdivision is not
motivated by physics, but by method: Most results are easiest to prove if you
restrict yourself to the regular domain and they can then be transferred to the
complementary domain using a~spin-flip mapping.

It is relatively simple to determine the LEC among all collinear states,
but these depend on the domain in which $\alpha_0$ lies. We
define two special collinear states that are used throughout the paper:
\begin{eqnarray}\label{defPsiUpUp}
  \bPsi_{\uparrow\ldots\uparrow}:&\; {\rm all}\;\;\psi_\mu=0
    \quad{\rm for}\;\; 0\le \mu <N,\\
  \label{defPsiUpDown}
  \bPsi_{\uparrow\ldots\downarrow}:&\; {\rm all}\;\;\psi_\mu=0
    \quad{\rm for}\;\;  0< \mu <N,\,\mu\neq \tau,
  \quad{\rm and}\;\;\psi_\tau=-\psi_0=\pi.
\end{eqnarray}
Here we have also fixed the values of~$\psi_0$ subject to the conventions
chosen below. Note that the definition~\eref{defPsiUpDown} possibly depends on
the choice of~$\tau$ if there are several FM coupling constants of minimal
strength.

\begin{prop}\label{propCollLEC}\mbox{~}

\begin{enumerate}
 \item For $0<\alpha_0<|\alpha_\tau|$ the state $\bPsi_{\uparrow\ldots\uparrow}$
   has a~lower energy than any other collinear state.
 \item For $|\alpha_\tau|<\alpha_0<\infty$ the state
   $\bPsi_{\uparrow\ldots\downarrow}$ has a~lower energy than any other
    collinear state (except those states $\bPsi_{\uparrow\ldots\downarrow}$
    resulting from a~different choice of~$\tau$).
 \item For $\alpha_0=|\alpha_\tau|$ the energy of
   $\bPsi_{\uparrow\ldots\uparrow}$ and of $\bPsi_{\uparrow\ldots\downarrow}$
   coincides and is lower than the energy of any other collinear state (except
   those states $\bPsi_{\uparrow\ldots\downarrow}$ resulting from a~different
   choice of~$\tau$).
\end{enumerate}
\end{prop}

\noindent\textbf{Proof.}~The state $\bPsi_{\uparrow\ldots\uparrow}$ satisfies
all bonds except the AFM bond $e_0=(0,1)$. Hence the only collinear state which
possibly assumes a~lower energy will be a~state satisfying all bonds, including
$e_0=(0,1)$, except another FM bond $e_\mu=(\mu,\mu+1)$. Further, $|\alpha_\mu|$
has to be minimal, which means $\mu=\tau$ (taking into account that $\tau$~need
not be unique). This implies that only $\bPsi_{\uparrow\ldots\downarrow}$ is
a~possible second candidate for an LEC among the collinear states. For the
energies we obtain
\begin{eqnarray}
\label{EPsiUpUp}
  E(\bPsi_{\uparrow\ldots\uparrow})
    &= \alpha_0-\sum_{\mu=1}^{N-1}|\alpha_\mu|, \\
  \label{EPsiUpDown}
  E(\bPsi_{\uparrow\ldots\downarrow})
    &=-\alpha_0-\sum_{\mu=1,\,\mu\neq\tau}^{N-1}|\alpha_\mu|+|\alpha_\tau|,
\end{eqnarray}
whence
\begin{equation}\label{EnDiff}
  E(\bPsi_{\uparrow\ldots\uparrow})- E(\bPsi_{\uparrow\ldots\downarrow})
    = 2(\alpha_0-|\alpha_\tau|),
\end{equation}
from which the remaining claims of the proposition follow.
\hfill$\square$
\bigskip

\section{Regular domain: $0<\alpha_0<|\alpha_\tau|$}\label{RegDom}

In the regular domain there are two prominent values of $\alpha_0$:
A critical point~$\alpha_{\rm (c)}$ which marks a~continuous phase
transition between the collinear and the coplanar ground state phase, and
a~point~$\alpha_{\rm (em)}$ where the energy, written as a~function
of~$\alpha_0$, assumes its maximum. Before we look at the theory behind this,
a~few preparations are necessary.

First consider the following

\begin{prop}\label{Pknull}\mbox{~}

\begin{enumerate}
  \item  Every coplanar LEC $\bPsi=(\psi_1, \ldots, \psi_{N-1})$ satisfies
    \begin{equation}\label{knull}
      \Psi + \psi_0=0,
    \end{equation}
    i.e., the integer~$k$ in \eref{AngCons} assumes the value $k=0$. Moreover,
    \begin{equation}\label{Psirange}
    0<\Psi<\pi.
    \end{equation}
  \item In the regular domain every LEC $\bPsi=(\psi_1, \ldots, \psi_{N-1})$
    satisfies
    \begin{equation}\label{PsiRC}
      0\le\psi_\mu\le\pi/2,\qquad {\rm for\;\; all}\quad 0<\mu<N.
    \end{equation}
\end{enumerate}
\end{prop}

The critical value $\alpha_{\rm (c)}$ mentioned above lies in the regular
domain, what is stated in the following
\begin{theor}\label{ThAlphC}
Assume a~classical spin systems as described in Theorem \ref{MTh} with all but
one couplings ferromagnetic, i.e., $\alpha_\mu<0$ for $0<\mu<N$ and
$\alpha_0>0$. It is also assumed that this unique antiferromagnetic exchange
integral is the weakest one, i.e.,
$0<\alpha_0<|\alpha_\tau|=\min_{0<\mu<N}|\alpha_\mu|$ and hence the spin system
will be in the regular domain. Then the lowest energy configuration of such
system is collinear if and only if $0<\alpha_0\le\alpha_{\rm (c)}$ with
\begin{equation}\label{AlphaC}
  \alpha_{\rm (c)}:=\left(\sum_{\mu=1}^{N-1}|\alpha_\mu|^{-1}\right)^{-1}
    < \left|\alpha_\tau \right|.
\end{equation}
\end{theor}

Before calculating the second prominent value~$\alpha_{\rm (em)}$ where the
energy~$E(\alpha_0)$ is maximal we recall the function
$f\colon[0,\pi]\to{\mathbb R}$ defined in the proof of Theorem~\ref{ThAlphC} by
\begin{equation}\label{deffrecall}
  f(\Psi)=
   \sum_{\mu=1}^{N-1}\arcsin\left(\frac{\alpha_0}{|\alpha_\mu|}\,
     \sin\Psi\right)  ,
\end{equation}
see \eref{deff}, and satisfying $\frac{\rmd^2 f}{\rmd\Psi^2}<0$ for all
$0<\Psi<\pi$. Every coplanar LEC~$\bPsi$ yields a~fixed point of~$f$,
$\Psi=f(\Psi)$. Whereas in the proof we argued that this equation cannot be
satisfied for a~non-trivial $\bPsi\neq{\mathbf 0}$ and
$0<\alpha_0\le\alpha_{\rm (c)}$ we now will use that a~coplanar LEC exists for
$\alpha_{\rm (c)}<\alpha_0<|\alpha_\tau|$ and hence the equation
$\Psi=f(\Psi)$ has at least one solution. By reconsidering \eref{fder10} we
conclude
\begin{equation}\label{fder11}
  \left.\frac{\rmd f}{\rmd\Psi}\right|_{\Psi=0}
    =\sum_{\mu=1}^{N-1 }\frac{\alpha_0}{|\alpha_\mu|}
    =\frac{\alpha_0}{\alpha_{\rm (c)}}> 1,
\end{equation}
for all $\alpha_{\rm (c)}<\alpha_0 <|\alpha_\tau|$.
In the following sums of the form $\sum_{\mu=1,\,\mu\neq N-\tau}^{N-1}$
will be abbreviated as ${\sum_\mu^\ast}$.
Then we have the following result:

\begin{prop}\label{Propunique}\mbox{~}

\begin{enumerate}
  \item For $\alpha_{\rm (c)}<\alpha_0 < |\alpha_\tau|$ the solution of
    $\Psi=f(\Psi)$ is unique and will be denoted by $\Psi(\alpha_0)$.
  \item The function $\alpha_0\mapsto \Psi(\alpha_0)$ is analytic for
    $\alpha_{\rm (c)} < \alpha_0  < |\alpha_\tau|$.
  \item If we keep $\alpha_0<\left| \alpha_\tau \right|$ fixed and consider $\Psi$ as a function of $\left|\alpha_{N-\tau}\right|$
then  $\left|\alpha_{N-\tau}\right|\mapsto \Psi\left(\left|\alpha_{N-\tau}\right|\right)$ is analytic for
\begin{equation}\label{conditionAlphaNminusTau}
 \left|\alpha_\tau\right| < \left|\alpha_{N-\tau} \right| <
 \left\{
 \begin{array}{r@{\quad:\quad}l}
  \left( \frac{1}{\alpha_0} - {{\sum_\mu^\ast}} \frac{1}{\left| \alpha_\mu \right|}\right)^{-1} &
   \frac{1}{\alpha_0} > {{\sum_\mu^\ast}}  \frac{1}{\left| \alpha_\mu \right|} \\
   \infty & \mbox{else}\;.
 \end{array}
 \right.
\end{equation}
 If $N-\tau = \tau \mbox{ mod } N$ then the lower bound of $\left|\alpha_{N-\tau} \right|$
 in \eref{conditionAlphaNminusTau} has to be chosen as $\alpha_0$.
\end{enumerate}
\end{prop}

This proposition implies that also the coplanar LEC~$\bPsi$ for
$\alpha_{\rm (c)}<\alpha_0<|\alpha_\tau|$ will be unique, since the
angles $\psi_\mu$, $0<\mu<N$, are determined by~$\Psi$ \textit{via\/}
\eref{MinCond1eval}. Hence a~discontinuous phase transition can be excluded for
the domain $\alpha_{\rm (c)}<\alpha_0<|\alpha_\tau|$.

The function $\alpha_0\mapsto \Psi(\alpha_0)$ cannot be extended analytically
across the critical value~$\alpha_{\rm (c)}$. In fact, there exists an
expansion of $\Psi(\alpha_0)$ in odd powers of
$\sqrt{\alpha_0-\alpha_{\rm (c)}}$ of which we will calculate the first term.
To this end we set
\begin{equation}\label{alpha0eps}
  \alpha_0 = \alpha_{\rm (c)}+\epsilon,\quad\quad \epsilon >0,
\end{equation}
and expand $f(\Psi)-\Psi$ into a~Taylor series at $\Psi=0$. Since~$f$ is an odd
function, the first two non-vanishing terms will be the linear and the cubic
term. This gives, after some calculations, the asymptotic form of the
non-vanishing solution of $f(\Psi)=\Psi$ as
\begin{equation}\label{PsiAsy}
  \Psi=\frac{\sqrt{\epsilon}}{\sqrt{A\,\alpha_{\rm (c)} }} +O(\epsilon^{3/2}),
\end{equation}
where
\begin{equation}\label{defA}
  A:=\frac{1}{6}\left(1-\alpha_{\rm (c)}^3
    \sum_{\mu=1}^{N-1}|\alpha_\mu|^{-3}\right).
\end{equation}
The corresponding asymptotic form of
$\beta(\alpha_0)=\alpha_0\sin\Psi(\alpha_0)$, see \eref{defBeta}, is given by
\begin{equation}\label{BetaAsy}
  \beta=\alpha_{\rm (c)}\, \frac{\sqrt{\epsilon}}{\sqrt{A\,\alpha_{\rm (c)} }}
    +O(\epsilon^{3/2}).
\end{equation}

Together with $\Psi$ also $\psi_\mu$, $0<\mu<N,$ and the energy~$E$ can be
written as analytic functions of $\alpha_0$ in the domain
$\alpha_{\rm (c)} <\alpha_0 <|\alpha_\tau|$. Any extremal value of~$E(\alpha_0)$
satisfies
\begin{eqnarray}
\nonumber
  0 &= \frac{\rmd E}{\rmd\alpha_0}\stackrel{\eref{EnbyPsi}}{=}
    \cos\Psi -\alpha_0\,\sin\Psi \frac{\rmd \Psi }{\rmd\alpha_0}
    +\sum_{\mu=1}^{N-1}|\alpha_\mu|
    \sin\psi_\mu \frac{\rmd\psi_\mu }{\rmd\alpha_0}\\
\label{EmaxCond2}
  &\stackrel{\eref{defBeta}}{=}  \cos\Psi -\beta
  \left(\frac{\rmd\Psi}{\rmd\alpha_0}-\sum_{\mu=1}^{N-1}
  \frac{\rmd\psi_\mu }{\rmd\alpha_0}\right) = \cos\Psi,
\end{eqnarray}
using $\Psi =\sum_{\mu=1}^{N-1}\psi_\mu$ according to \eref{Psi0}. The extremum
of~$E(\alpha_0)$ is hence assumed for $\Psi=\pi/2$, the only zero of $\cos\Psi$
in the interval $0\le \Psi\le \pi$. The corresponding value of~$\alpha_0$ will
be denoted by $\alpha_{\rm (em)}$ and is, according to~\eref{deff}, obtained by
the solution of
\begin{equation}\label{alphaEM}
  \frac{\pi}{2}=\sum_{\mu=1}^{N-1}\arcsin\frac{\alpha_{\rm (em)}}
    {|\alpha_\mu|}.
\end{equation}
Note that the function
$x\mapsto\sum_{\mu=1}^{N-1}\arcsin(x/|\alpha_\mu|)$
is strictly monotonically increasing for $0\le x\le|\alpha_\tau|$ and assumes
a~value greater than $\pi/2$ at $x=|\alpha_\tau|$. Hence the solution
$x=\alpha_{\rm (em)}$ of \eref{alphaEM} is unique and satisfies
\begin{equation}\label{alphaENalphaC}
  \alpha_{\rm (em)}<|\alpha_\tau|.
\end{equation}
The extremum of $E(\alpha_0)$ is in fact a~maximum, as can be confirmed by
calculating the second derivative. Without going into the details we present the
result in the form
\begin{equation}\label{Emaxder}
  E(\alpha_0)\simeq E_{\rm max}- e_2
    \left(\alpha_0-\alpha_{\rm (em)} \right)^2,
\end{equation}
where
\begin{equation}\label{EmaxE2}
  E_{\rm max}=-\sum_{\mu=1}^{N-1}\sqrt{\alpha_\mu^2-\alpha_{\rm (em)}^2}
\end{equation}
and
\[
  e_2:=\frac{1}{2}\sum_{\mu=1}^{N-1}
    \left(\alpha_\mu^2-\alpha_{\rm (em)}^2\right)^{-1/2}>0.
\]
The arguments of the square roots are positive due to \eref{alphaENalphaC} and
$|\alpha_\tau|\le|\alpha_\mu|$ for all $0<\mu<N$.

\section{Complementary domain: $|\alpha_\tau|<\alpha_0<\infty$}\label{ComDom}
\begin{figure}[tb]
\begin{center}
  \includegraphics[width=15cm]{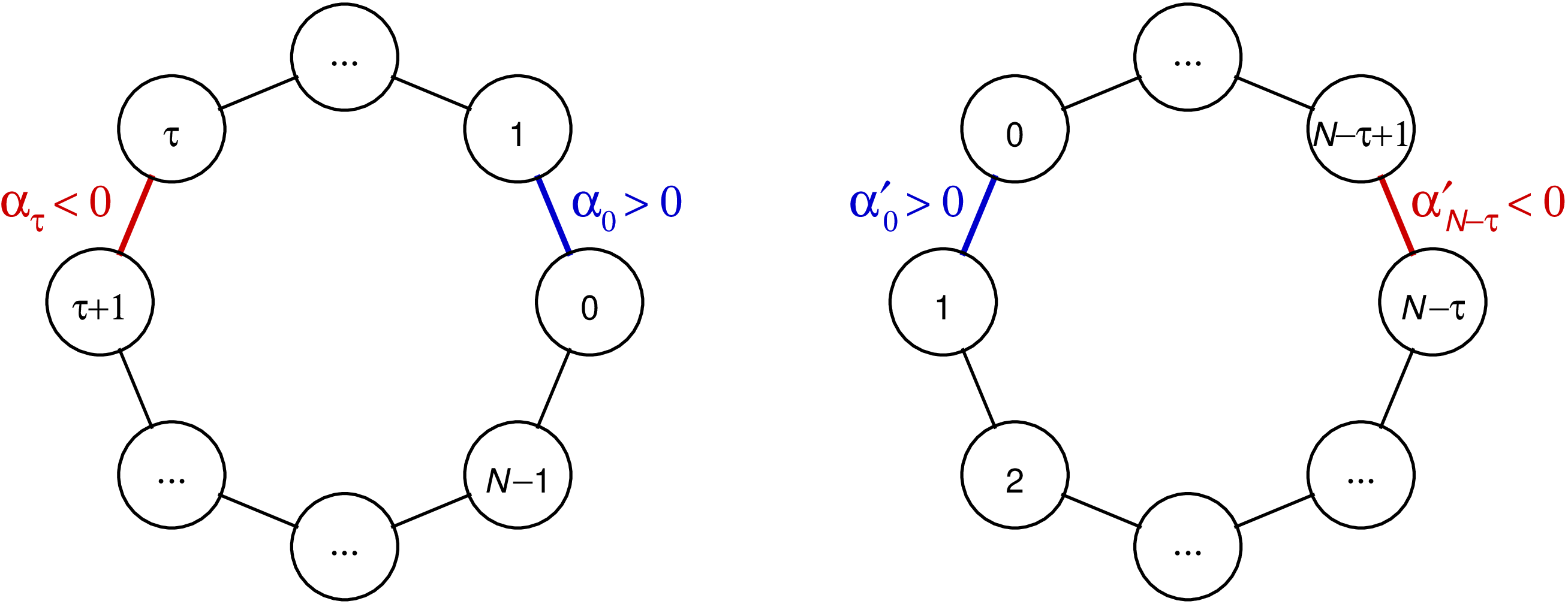}
\end{center}
\caption{Left panel: Sketch of a polygon with AFM coupling $\alpha_0>0$ and the
weakest FM coupling $\alpha_\tau<0$. Right panel: The transformed polygon after
a~flip-transformation and a~renumbering of spin sites.\label{RR}}
\end{figure}

For $\alpha_0>|\alpha_\tau|$ we cannot apply the theory presented in
\sref{RegDom}. One way out is to flip the spins of number $1,\ldots,\tau$ and
change the signs of the two coupling constants~$\alpha_0$ and~$\alpha_\tau$
in order to obtain a~polygon that lies in the regular domain and thus fulfills
the conditions of the previous theory. It is advisable to renumber the resulting
polygon so that it corresponds to the previous conventions, see~\fref{RR}.

We compile the resulting transformations, whereby the variables of the
transformed polygon are marked with an apostrophe:
\numparts\begin{eqnarray}
\label{transform1}
  \alpha_0' &= -\alpha_\tau >0, \\
\label{transform2}
  \alpha_{N-\tau}' &= -\alpha_0 <0, \\
\label{transform3}
  \alpha_{\mu}' &= \alpha_{\mu+\tau}, \quad
    0<\mu<N,\; \mu\neq N-\tau,\\
\label{transform4}
  \psi_{0}' &= \psi_\tau-\pi<0, \quad {\rm such \; that\;\;}
    \Psi'=- \psi_{0}' =\pi- \psi_\tau>0,\\
\label{transform5}
  \psi_{N-\tau}'&=\psi_0+\pi=\pi-\Psi,\\
\label{transform6}
   \psi_{\mu}'&=\psi_{\mu+\tau},\quad
     0<\mu<N,\; \mu\neq N-\tau.
\end{eqnarray}\endnumparts

The transformation \eref{transform1}--\eref{transform6}, shortly called the
`flip-transformation' $({\mathcal P},\bPsi)\mapsto({\mathcal P}',\bPsi')$,
where ${\mathcal P}$ symbolizes a~polygon and~$\bPsi$ its LEC, maps the domain
where the theory of \sref{RegDom} cannot be applied into the domain where it
works. The results for $({\mathcal P}',\bPsi')$ can then be re-translated in
terms of  $({\mathcal P},\bPsi)$. For example, if $\alpha'_0>\alpha'_{\rm (c)}$
then the LEC will be coplanar.

In principle, this concludes the description of the LEC for all values of the
exchange integrals $\alpha_\mu$, $0\le \mu<N$. Nevertheless, it will be
interesting to investigate what happens if the one-parameter family of
polygons~${\mathcal P}(\alpha_0)$ meets points that are mapped under the
flip-transformation onto critical points with $\alpha'_0=\alpha'_{\rm (c)}$
or $\alpha'_0=\alpha'_{\rm (em)}$. Note that for the flip-transformed
one-parameter family of polygons $\alpha'_\tau$ is variable and $\alpha'_0$
is kept fixed.

We start with the second case. At $\alpha'_0=\alpha'_{\rm (em)}$ we have
$\sin\Psi'=1$ and hence $\beta=\beta'= \alpha'_0 \sin\Psi'$ is maximal since
$\alpha'_0\stackrel{\eref{transform1}}{=}\left|\alpha_\tau \right|$ is kept
fixed. We will re-translate the condition $\alpha'_0=\alpha'_{\rm (em)}$ into
a~condition for $\alpha_0$ and consider
\begin{eqnarray}
\nonumber
  \frac{\pi}{2} &\stackrel{\eref{alphaEM}}{=} \sum_{\mu=1}^{N-1}
    \arcsin\frac{\alpha'_0}{|\alpha'_\mu|} \\
\nonumber
  &= \sum_{\mu=0,\,\mu\neq N-\tau}^{N-1}\arcsin
  \frac{\alpha'_0}{|\alpha'_\mu|}
    -\arcsin\frac{\alpha'_0}{\alpha'_0}
    +\arcsin\frac{\alpha'_0}{|\alpha'_{N-\tau}|}\\
\label{retrans3}
  &\stackrel{\eref{transform1}\eref{transform2}}{=}
    \sum_{\mu=1}^{N-1}\arcsin\frac{|\alpha_\tau|}{|\alpha_\mu|}
    -\frac{\pi}{2}+\arcsin\frac{|\alpha_\tau|}{\alpha_{0}}.
\end{eqnarray}
Solving \eref{retrans3} for $\alpha_0$ gives
\begin{eqnarray}\label{retrans4}
  \alpha_0&= |\alpha_\tau|\left(\sin \left( \sum_{\mu=1}^{N-1}
  \arcsin\frac{|\alpha_\tau|} {|\alpha_\mu|} -\frac{\pi}{2}\right)\right)^{-1}\\
  \label{retrans5}
  &= |\alpha_\tau|\left(\cos \sum_{\mu=1,\,\mu\neq \tau}^{N-1}
   \arcsin\frac{|\alpha_\tau|}{|\alpha_\mu|} \right)^{-1}
   =: \alpha_{\rm (bm)}.
\end{eqnarray}
It cannot be assured that $0< \alpha_{\rm (bm)}<\infty$ for all possible values
of $\alpha_\mu$, $0<\mu<N$, and hence that $\beta(\alpha_0)$ assumes the
maximum. In \sref{Ex} we will encounter examples where $\alpha_{\rm (bm)}<0$
and consequently $\beta(\alpha_0)$ is monotonically increasing without reaching
the maximal value. If the maximum~$\beta_{\rm max}$ is
reached then we have
\begin{equation}\label{betamax}
\beta_{\rm max} = \alpha'_0 \sin\Psi'=|\alpha_\tau|.
\end{equation}

Next we will re-translate the condition $\alpha'_0=\alpha'_{\rm (c)}$ into
a~condition for $\alpha_0$ and consider
\begin{equation}\label{alphapc}
  \alpha'_0=\alpha'_{\rm (c)}\stackrel{\eref{AlphaC}}{=}
    \left(\sum_{\mu=1}^{N-1}|\alpha'_{\mu}|^{-1} \right)^{-1}.
\end{equation}
Due to \eref{transform1} and \eref{transform2} this is equivalent to
\begin{eqnarray}
\nonumber
  |\alpha_\tau|^{-1} &= \sum_{\mu=0,\,\mu\neq N-\tau}^{N-1}
    |\alpha'_{\mu}|^{-1}
    -(\alpha'_0)^{-1}+|\alpha'_{N-\tau}|^{-1}\\
\label{retrans6}
   &=  \sum_{\mu=1}^{N-1}
    |\alpha_{\mu}|^{-1}-|\alpha_\tau|^{-1}
    +\alpha_{0}^{-1}=\alpha_{\rm (c)}^{-1}
    -|\alpha_\tau|^{-1}+ \alpha_{0}^{-1},
\end{eqnarray}
and further to
\begin{equation}\label{retrans7}
  \alpha_0=\left(2|\alpha_\tau|^{-1}
    -\alpha_{\rm (c)}^{-1} \right)^{-1}=: \alpha_{\rm (c')}.
\end{equation}
Also in this case examples show that $0<\alpha_{\rm (c')}<\infty$ need not be
satisfied for all possible values of $\alpha_\mu$, $0<\mu<N$, and hence that
a~transition to a~collinear LEC $\bPsi_{\uparrow\ldots\downarrow}$
is possible but not necessary. Introducing
\begin{equation}\label{Alphf}
  \alpha_{\rm (d)}^{-1}:=\sum_{\mu=1,\,\mu\neq\tau}^{N-1}|\alpha_\mu|^{-1}
    =\alpha_{\rm (c)}^{-1}-|\alpha_\tau|^{-1}
\end{equation}
the definition \eref{retrans7} can be rewritten as
\begin{equation}\label{retrans7a}
  \alpha_{\rm (c')}^{-1}=|\alpha_\tau|^{-1}-\alpha_{\rm (d)}^{-1}.
\end{equation}
Hence, $0<\alpha_{\rm (c')}<\infty$ iff $|\alpha_\tau|<\alpha_{\rm (d)}$.
In this case the second phase transition to the collinear LEC
$\bPsi_{\uparrow\ldots\downarrow}$ will occur where
$\bPsi_{\uparrow\ldots\downarrow}$ is the flip-transform of the fully aligned
collinear configuration $\bPsi_{\uparrow\ldots\uparrow}$. At this transition
$\Psi(\alpha_0)$ will approach the value~$\pi$ asymptotically similar as in
\eref{PsiAsy}:
\begin{equation}\label{Psi1Asy}
  \Psi=\pi-\frac{\alpha_\tau^2}{\alpha_{\rm (c')}^2}
    \sqrt{\frac{\epsilon}{B\,|\alpha_\tau|}}+O(\epsilon^{3/2}),
\end{equation}
where
\begin{equation}\label{defB}
  B:= \frac{1}{6}\left[1-  |\alpha_\tau|^3
    \left(\sum_{\mu=1}^{N-1}|\alpha_\mu|^{-3}
    -|\alpha_\tau|^{-3}+\alpha_{\rm (c')}^{-3}\right) \right]
\end{equation}
and $\epsilon:= \alpha_{(c')}-\alpha_0>0$.

Analogously, in the case of positive $\alpha_{\rm (c')}$, $\beta(\alpha_0)$ will
reach its limit value~$0$ for $\alpha_0\to\alpha_{\rm (c')}$ asymptotically
in the form
\begin{equation}\label{betaAsy1}
  \beta=\frac{\alpha_\tau^2}{\alpha_{\rm (c')}}
    \sqrt{\frac{\epsilon}{B\,|\alpha_\tau|}}+O(\epsilon^{3/2}).
\end{equation}

\section{Phase diagram}\label{PD}

\begin{figure}[tb]
\begin{center}
 \includegraphics[width=15cm]{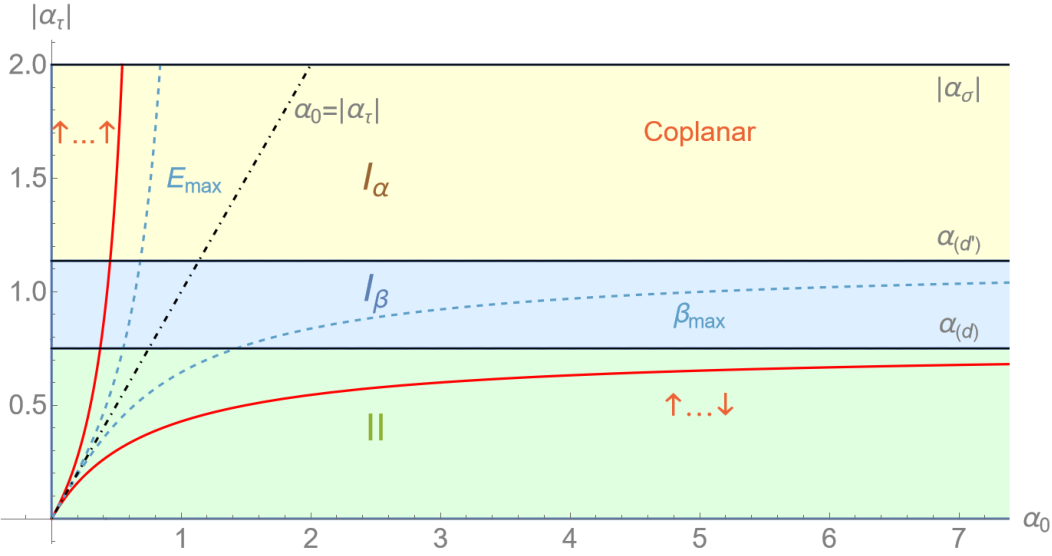}
\end{center}
\caption{
Phase diagram of an $N=6$ polygon with FM bonds given by \eref{hexadata}. The
coordinates are the AFM bond~$\alpha_0$ and the weakest FM
bond~$|\alpha_\tau|=|\alpha_1|$. The (red) solid curves separate the coplanar
and the two collinear phases. The (blue) dashed curves indicate the points where
the energy or the parameter~$\beta$ assume their maxima w.r.t.~$\alpha_0$. The
dash-dotted line $\alpha_0=|\alpha_\tau|$ separates the `regular domain' from
the `complementary domain'. The three horizontal layers
(indicated by the colors light green, light blue, light yellow)
correspond to the case distinction between type $II$, $I_\beta$ and $I_\alpha$
of Theorem \ref{TPD}. Note the symmetry of the phase diagram w.~r.~t.~the reflection
$\alpha_0 \leftrightarrow |\alpha_\tau|$. This reflection maps the curve denoted by
$E_{\rm max}$, characterized by $\Psi=\pi/2$, onto the curve denoted by $\beta_{\rm max}$,
and characterized by $\Psi'=\pi/2$,
}
\label{FIGPD}
\end{figure}

We have seen in the previous section that there are three qualitative different
possibilities for the one-parameter family of polygons/LECs which will be denoted
by $\alpha_0\mapsto({\mathcal P}(\alpha_0),\bPsi(\alpha_0))$ and that these
possibilities depend on the value of the weakest FM bond $\alpha_\tau$. In this
section we will investigate these dependencies more systematically and to this
end consider not only~$\alpha_0$ but also~$\left|\alpha_\tau\right|$ as variable
and the remaining FM bonds as fixed. This leads to a $2$-dimensional phase
diagram showing boundaries between the collinear and coplanar LEC-phases and
further curves defined by maximal energy or maximal~$\beta$, see \fref{FIGPD}.

It will be convenient to choose the (second) weakest FM bond~$\alpha_\sigma$
among the remaining FM bonds~$\alpha_\mu$, $0<\mu<N$, $\mu\neq \tau$, and to
restrict the domain of~$|\alpha_\tau|$ to the interval~$(0,|\alpha_\sigma|]$.
Moreover, we will say that the  one-parameter family of polygons/LECs, shortly
referred to as the `$\alpha_0$-family', is of `type~I' iff there occurs only
one phase transition, otherwise it is said to be of `type~II'. If, in the first
case, the function $\alpha_0\mapsto \beta(\alpha_0)$ has the maximum then the
$\alpha_0$-family will be said to be of `type~I$_\beta$', otherwise of
type~`I$_\alpha$'. Sums of the form $\sum_{\mu=1,\,\mu\neq \tau}^{N-1}$ will be
abbreviated as~$\sum_\mu'$. We recall the definition~\eref{Alphf}
\begin{equation}\label{defAlphaF}
  \alpha_{\rm(d)}^{-1}= \left.\sum\right.'_\mu \left|\alpha_\mu \right|^{-1},
\end{equation}
and introduce another characteristic value $ \alpha_{\rm(d')}$ by a~solution of
the equation
\begin{equation}\label{defAlphaG}
  \left.\sum\right.'_\mu \arcsin \frac{\alpha_{\rm(d')}}{|\alpha_\mu|}
    =\frac{\pi}{2}.
\end{equation}
Then the relationship between the type of the $\alpha_0$-family and
the value of $\left|\alpha_\tau\right|$ is summarized in the following
\begin{theor}\label{TPD}
If $N>3$ then:
  \begin{enumerate}
    \item The equation \eref{defAlphaG} has the unique solution
    $\alpha_{\rm(d')}$ satisfying
    \begin{equation}\label{order}
      0< \alpha_{\rm(d)} <\alpha_{\rm(d')}< |\alpha_\sigma|.
    \end{equation}
    \item $0<|\alpha_\tau|< \alpha_{\rm(d)}$ iff the
      $\alpha_0$-family is of type II,
    \item $\alpha_{\rm(d)}\le|\alpha_\tau|<\alpha_{\rm(d')}$ iff the
    $\alpha_0$-family is of type I$_\beta$,
  \item $\alpha_{\rm(d')}\le|\alpha_\tau|\le |\alpha_\sigma|$
    iff the $\alpha_0$-family is of type  I$_\alpha$.
  \end{enumerate}
\end{theor}
Due to the assumption $|\alpha_\tau|\le|\alpha_\sigma|$ the enumeration
(ii)--(iv) is a~complete case distinction and hence we are sure that
for the $\alpha_0$-family only the three types II, I$_\alpha$ and I$_\beta$
will occur.

The 3-spin system has to be excluded since in this case
$\arcsin(\alpha_{\rm(d')}/|\alpha_\sigma|)=\pi/2$, so
\begin{equation}\label{Alphas}
  \alpha_{\rm(d')}=|\alpha_\sigma|\stackrel{\eref{Alphf}}{=}\alpha_{\rm(d)}
\end{equation}
and the type I$_\alpha$ is observed for $|\alpha_\tau|=|\alpha_\sigma|$
(i.e., $\alpha_1=\alpha_2$) only, whereas the type I$_\beta$ is absent
(see also \sref{Ex}).

We conclude this section by considering the limit $\alpha_0\to\infty$
of certain quantities in the case that the $\alpha_0$-family is of type~I.

Obviously, $\Psi\to\pi$ for $\alpha_0\to\infty$. According to the flip
transformation \eref{transform1}--\eref{transform6} we have
\begin{equation}\label{betaprime}
 \beta=\beta'=|\alpha_0'| \sin \Psi' =|\alpha_\tau| \sin \psi_\tau.
\end{equation}
It therefore makes sense to examine the limit value $\Psi_\infty'$ of $\Psi'$
because this results in the limit value
$\beta_\infty=|\alpha_\tau| \sin \Psi_\infty'$ of $\beta$. We obtain the
following result.
\begin{prop}\label{PropLimBeta}
$\Psi_\infty'$ is the solution of one of the following equations,
depending on the type of the $\alpha_0$-family.
\begin{eqnarray}
\label{PsiprimeLimitalpha}
I_\alpha\colon\qquad  \pi &=& \sum_{\mu=1}^{N-1}\arcsin
   \left( \frac{|\alpha_\tau|}{|\alpha_\mu|}\sin \Psi_\infty'\right),\\
\label{PsiprimeLimitbeta}
I_\beta\colon\qquad \Psi_\infty' &=& \frac{1}{2}\sum_{\mu=1}^{N-1}\arcsin
   \left(\frac{|\alpha_\tau|}{|\alpha_\mu|}\sin \Psi_\infty'\right).
\end{eqnarray}
In both cases we have
\begin{equation}
 \label{betaLimit}
   \beta_\infty = |\alpha_\tau| \sin \Psi_\infty'.
\end{equation}
\end{prop}

\section{Examples}\label{Ex}
In this section we present some results for different systems in order to illustrate
the most important and the most characteristic features. However, we will start with
a~brief discussion of the methods for determining the required values. In general,
there are three types of problems. The easiest is to identify some values
of the parameter~$\alpha_0$, when a~simple substitution will suffice
[cf.\ \eref{AlphaC}]. Determination of the parameters of the asymptotic solutions,
like in \eref{defA}, also belongs to this class. The second type consists transcendent
equations, such as \eref{alphaEM}, which are solved
numerically, sometimes with the use of symbolic algebra packages.

\begin{figure}
\begin{center}
  \includegraphics[width=10cm]{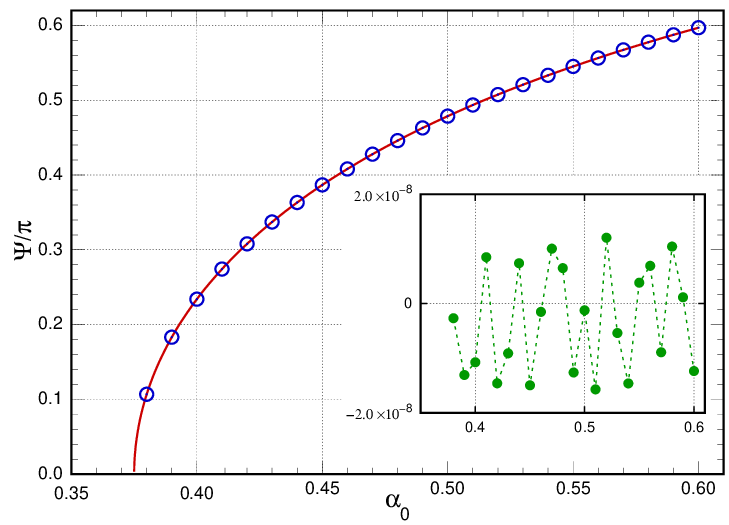}
\end{center}
\caption{The angle $\Psi=-\psi_0$ in the units of~$\pi$ as a~function of~$\alpha_0$
for $N=3$, $\alpha_1=-0.6$, and $\alpha_2=-1$ determined by the iterative
minimization method ($\Psi_{\rm IM}$, empty circles) and by the secant method
($\Psi_{\rm SM}$, solid line). In the inset the difference
$\Psi_{\rm IM}-\Psi_{\rm SM}$ is presented.\label{Cmp}}
\end{figure}

The most important data are obtained determining a~fixed point $\Psi=f(\Psi)$ of
the function $f\colon [0,\pi] \to {\mathbb R}$ defined in the proof of
Theorem~\ref{ThAlphC} [see also \eref{deffrecall}]. Hence, the root of
the function $\Psi-f(\Psi)$ has to be found. Standard methods (e.g.\ the secant one)
sometimes fail to work properly near the crtical point~$\alpha_{\rm (c)}$ since, see
\eref{PsiAsy},
$\lim_{\alpha_0\to\alpha_{\rm (c)}^+}\frac{\rmd\Psi}{\rmd\alpha_0}=\infty$.
Morover, the formula \eref{deffrecall} is valid in the interval
$\left(\alpha_{\rm (c)},|\alpha_\tau|\right)$ only. To cover the interval
$\left(|\alpha_\tau|,\alpha_{\rm (c')}\right)$ the spin-flip transformation has to
be performed. Therefore, the \emph{iterative minimization\/} method, shortly
presented in \cite{Schm17a} (see \cite{Baez2018} for detailed description), is mainly
used do determine the angle sequence~$\bPsi_0=(\psi_0,\psi_1,\ldots,\psi_{N-1})$.
Note that this methods gives all angles (and the system energy), though it is
enough to find the angle~$\Psi$, since the remaining angles and the Lagrange
parameter~$\beta$ can be found from proposition~\ref{propbeta}. Results obtained
by the secant and by the iterative minimization methods are compared in \fref{Cmp}
for $N=3$ with the equations $|\alpha_1|=0.6$ and $|\alpha_2|=1$.

\subsection{The special case: $N=3$}
The three-spin system is a~special case in two respects: (i)~the type~$I_\beta$ is
not observed (see Theorem~\ref{TPD}) and (ii)~some equations can be solved strictly.
Moreover, this system was considered in many papers (see, e.g.,
\cite{Schm03,Flor16,Schm17a,Dmit19,Schm22}), so some results may be easily compared.
Without loss of generality we may assume that $\alpha_2=-1$ and $0<|\alpha_1|\le 1$,
i.e., $\tau=1$.

\begin{figure}
\begin{center}
 \includegraphics[width=15cm]{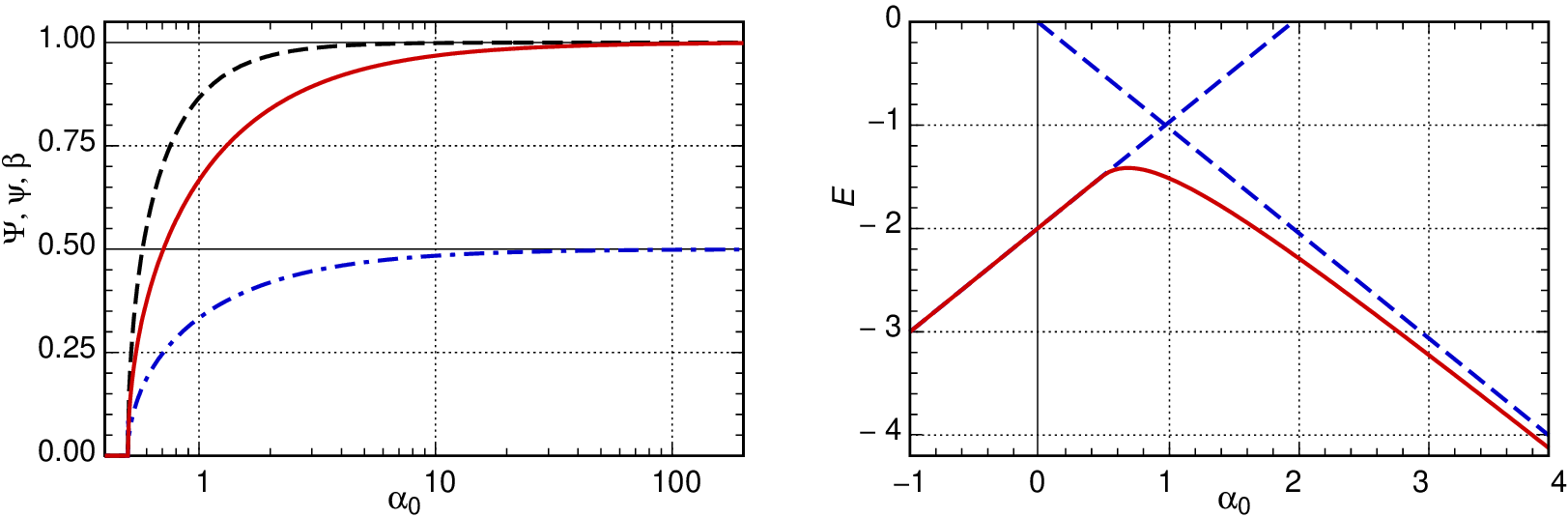}
\end{center}
\caption{The three-spin system with $\alpha_1=\alpha_2=-1$. Left panel: The angles (in
the units of~$\pi$) $\Psi=-\psi_0$ (red solid line) and $\psi=\psi_1=\psi_2$ (blue
dash-dotted line) with the parameter~$\beta$ (black dashed line) as functions
of~$\alpha_0$ (log-linear scale). Right panel: The LEC energy $E(\alpha_0)$ (red solid
line). The blue dashed lines show the energy of $\bPsi_{\uparrow\ldots\uparrow}$ (left
line) and $\bPsi_{\uparrow\ldots\downarrow}$ (right line) configurations; the latter
one is an asymptote to $E(\alpha_0)$ for $\alpha_0\to\infty$.\label{N3Alf1}}
\end{figure}

From \eref{AlphaC} we have
\begin{equation}\label{AlphaC3}
  \alpha_{\rm (c)}=\frac{|\alpha_1|}{1+|\alpha_1|}.
\end{equation}
It is also easy to determine the energy maximum. Since in this case
$\Psi=\pi/2$, then $\psi_2^{\rm (em)}=\pi/2-\psi_1^{\rm (em)}$ and, see
\eref{defBeta},
\begin{equation}\label{BetaRel}
  \alpha_{\rm (em)}=|\alpha_1|\sin\psi_1^{\rm (em)}
    =\sin\left(\pi/2-\psi_1^{\rm (em)}\right)=\cos\psi_1^{\rm (em)}.
\end{equation}
Therefore, $\cot\psi_1^{\rm (em)}=|\alpha_1|$ and
\begin{equation}\label{Alph0EM}
  \alpha_{\rm (em)}\stackrel{\eref{BetaRel}}{=}
  \cos\psi_1^{\rm (em)}=\frac{|\alpha_1|}{\sqrt{1+\alpha_1^2}}.
\end{equation}
It is easy to show, cf.\ \eref{EmaxE2}, that
\begin{equation}\label{EMax3}
  E_{\rm max}=\alpha_1\cos\psi_1^{\rm (em)}-\sin\psi_1^{\rm (em)}
   =-\sqrt{1+\alpha_1^2}.
\end{equation}

For $0<|\alpha_\tau|<1$ a~system is in the type~II and the second phase
transition takes place at
\begin{equation}\label{AlphaE3}
  \alpha_{\rm (c')}=\frac{|\alpha_\tau|}{1-|\alpha_\tau|}.
\end{equation}
The function $\beta(\alpha_0)$ reaches its maximume when $\psi_1=\pi/2$, so
from~\eref{defBeta} we have
\begin{equation}\label{BetaMRel}
  \alpha_{\rm (bm)}\sin\left(\pi/2+\psi_2^{\rm (bm)}\right)
  =\alpha_{\rm (bm)}\cos\psi_2^{\rm (bm)}=|\alpha_1|
    =\sin\psi_2^{\rm (bm)},
\end{equation}
hence we obtain
\begin{equation}\label{BMax3}
   \alpha_{\rm (bm)} = \frac{|\alpha_1|}{\sqrt{1-\alpha_1^2}}.
\end{equation}

In the very special case $\alpha_1=\alpha_2=-1$ we have $\psi_1=\psi_2=\psi$, so
\[
  \alpha_0\sin(2\psi)=\sin\psi
\]
and therefore for $\alpha_0\ge\alpha_{\rm (c)}=1/2$
\begin{equation}\label{Cospsi}
  2\alpha_0\cos\psi=1.
\end{equation}
The maximum of energy $E_{\rm max}=-\sqrt{2}$ is reached at $\alpha_{\rm (em)}=1/\sqrt{2}$
and the parameter~$\beta$ does not reach the maximum, so this system is in the
type~I$_\alpha$. If $\alpha_0\to\infty$ then $\Psi\to\pi$, so $\psi\to\pi/2$. Hence,
$\beta_\infty=|\alpha_1|=1$. It should be emphasized that the convergence of these angles
and the parameter~$\beta$ to their limit values is rather slow, see \fref{N3Alf1}.

\subsection{A hexagon with a single AFM bond}

\begin{table}[bt]
\caption{\label{HexParam}
Basic parameters of a~hexagon with FM bonds given by \eref{hexadata} for three
values of $|\alpha_\tau|$ determining different types of the $\alpha_0$-family.}
\begin{tabular}{@{}*{9}{l}}
\hline\hline\hline
Type & $|\alpha_\tau|$ & $\alpha_{\rm (c)}$ & $\alpha_{\rm (c')}$
   & $\alpha_{\rm (em)}$ & $E_{\rm max}$ & $\alpha_{\rm (bm)}$
   &  $\pi-\Psi'_\infty$ [rad]& $\beta_\infty/|\alpha_\tau|$ \\
   \hline
 II         & 1/3 & 3/13 & 3/5 & 0.305866 & $-14.070$ & 0.369425
    & $\pi^{\rm a}$& $0.0^{\rm a}$\cr\ms
 I$_\beta$  & 4/5 &  12/31 & n.a. & 0.573571 & $-14.336$ & 1.713753
  & 2.495577 & 0.60201 \\
 I$_\alpha$ & 3/2 & 1/2  & n.a. & 0.764801 & $-14.891$ & n.a.
  & 1.180456  & 0.92478\\
   \hline\hline\hline
\end{tabular}

$^{\rm a}$ For type II the collinear LEC
$\bPsi_{\uparrow\ldots\downarrow}$ is realized for $\alpha_0\ge \alpha_{\rm (c')}$.
\end{table}

\begin{figure}[tb]
\begin{center}
  \includegraphics[width=15cm]{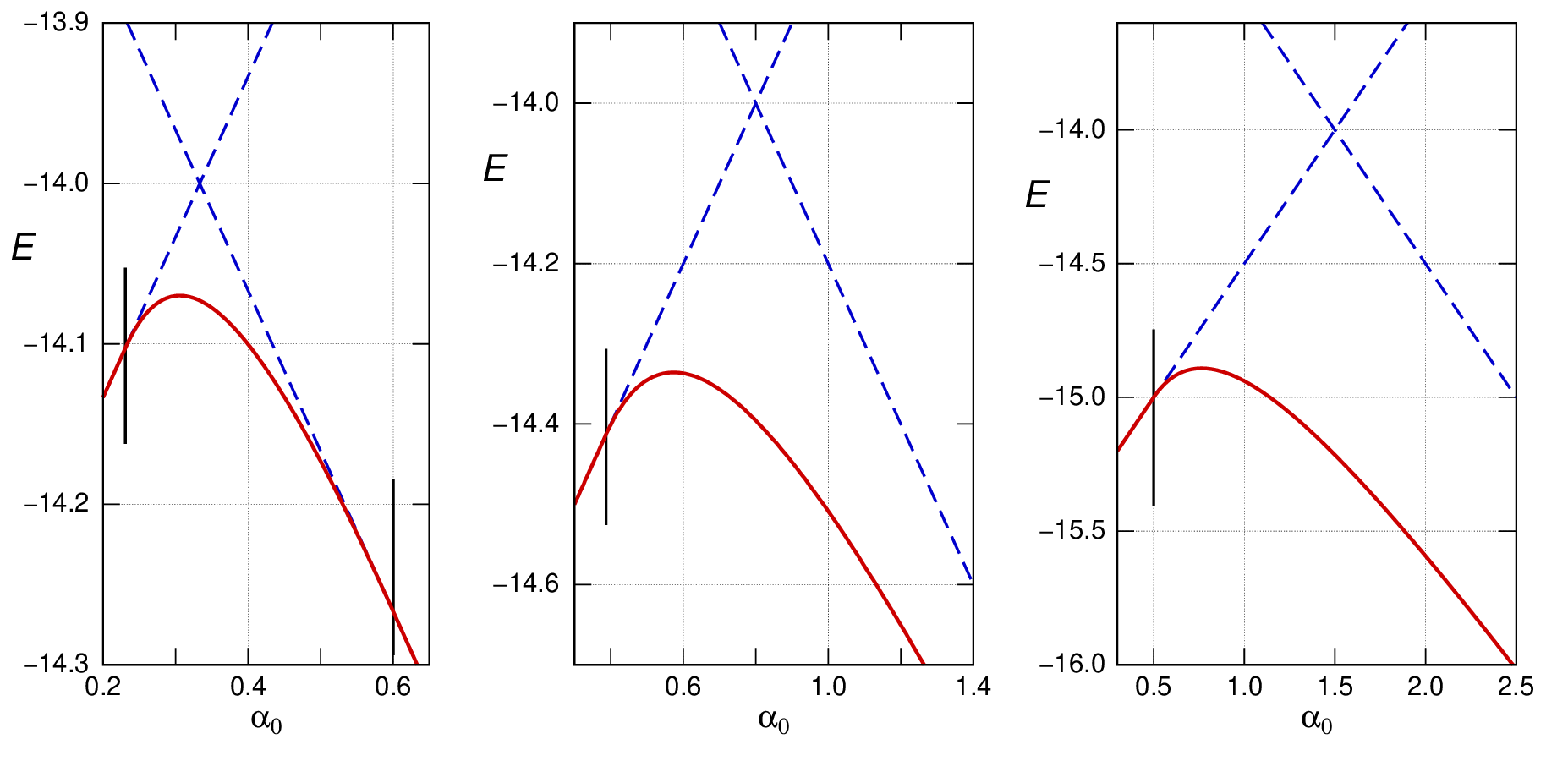}
\end{center}
\caption{The LEC energy $E$ for a~hexagon with FM bonds given by \eref{hexadata}
as a~function of~$\alpha_0$ (red solid curves). The blue dashed lines show the energy
of $\bPsi_{\uparrow\ldots\uparrow}$ (left line) and
$\bPsi_{\uparrow\ldots\downarrow}$ (right line) configurations; solid vertical
lines indicate the value of~$\alpha_{\rm (c)}$ (all panels) and
$\alpha_{\rm (c')}$ (left panel only). Left panel: $|\alpha_\tau|=1/3$ (type~II);
middle panel: $|\alpha_\tau|=4/5$ (type~I$_\beta$); right panel:
$|\alpha_\tau|=3/2$ (type~I$_\alpha$). \label{n6ener}}
\end{figure}

As a~more complex example we consider a~hexagon with the following fixed FM bonds
\begin{equation}\label{hexadata}
 \alpha_2=-2,\; \alpha_3=\alpha_4=-3,\; \alpha_5=-6.
\end{equation}
Therefore, the boundaries of the different types of the $\alpha_0$-family
in theorem \ref{TPD} are given as follows (see also the phase diagram in \fref{FIGPD})
\[
  \alpha_{\rm (d)}\stackrel{\eref{Alphf}}{=}3/4,\qquad
  \alpha_{\rm (d')}\stackrel{\eref{defAlphaG}}{\approx} 1.135685,\qquad
  |\alpha_\sigma|=2.
\]

\begin{figure}[tb]
\begin{center}
  \includegraphics[width=15cm]{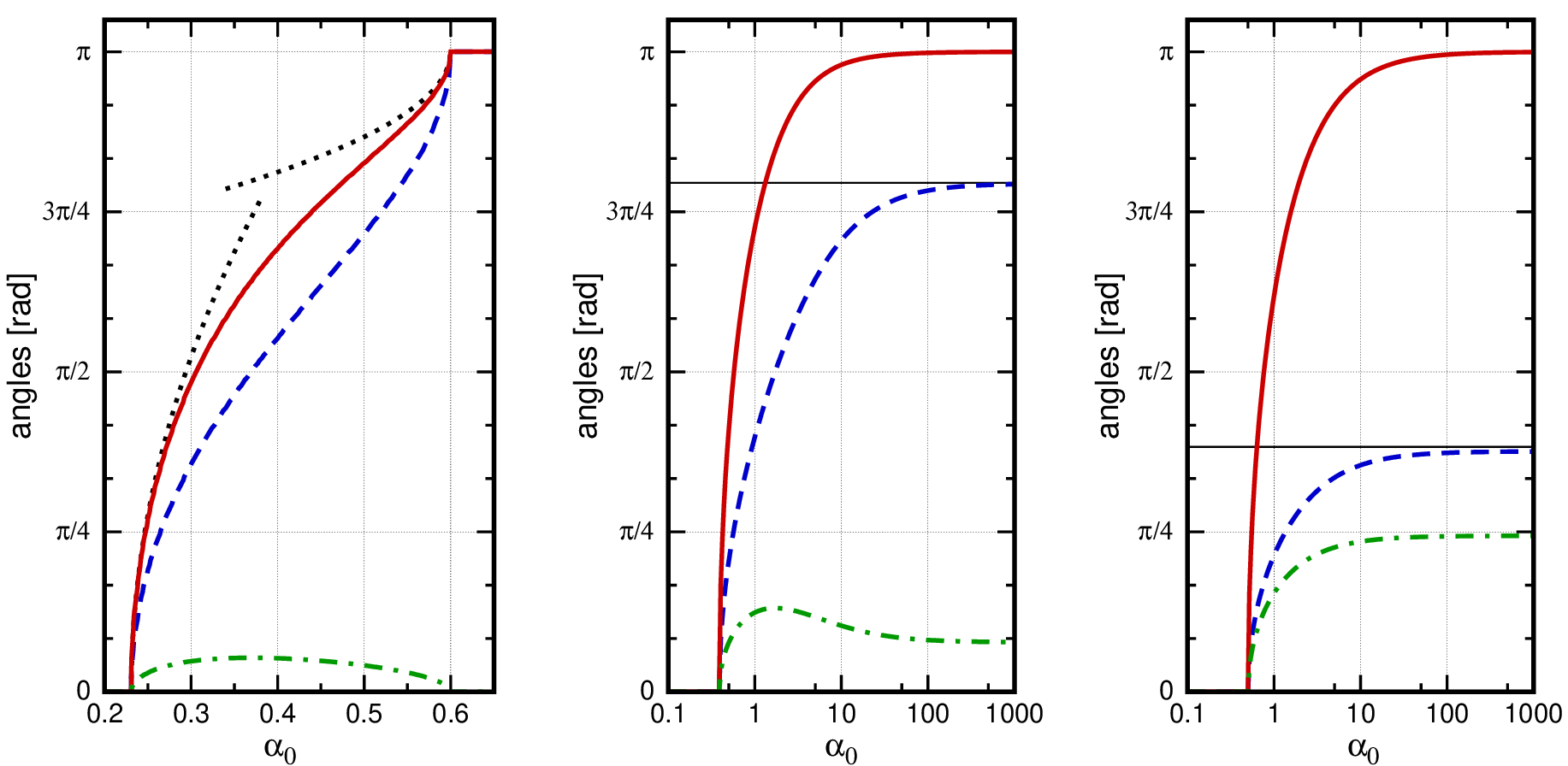}
\end{center}
\caption{The angles $\Psi=-\psi_0$ (red solid curves), $\psi_\tau\equiv\psi_1$ (blue
dashed), and $\psi_\sigma\equiv\psi_2$ (green dash-dotted curve) for a~hexagon with
FM~bonds given by \eref{hexadata}
as a~function of~$\alpha_0$; the angles $\psi_\mu$ for $\mu>2$ behave similarly as~$\psi_2$
with their values satisfying~\eref{defBeta}. In the left panel ($|\alpha_\tau|=1/3$,
 type~II) the asymptotic behaviour is indicated by dotted curves, see \eref{PsiAsy} and
\eref{Psi1Asy}. In types I$_\beta$ (middle panel, $|\alpha_\tau|=4/5$) and
I$_\alpha$ (right panel, $|\alpha_\tau|=3/2$) the angle~$\psi_1$  tends to its
limit values~$\pi-\Psi'_\infty$, see \eref{transform4}.
These values are given in \tref{HexParam}.
For the middle and the right panel a log-linear scale is used.
\label{n6angl}}
\end{figure}

\begin{figure}[tb]
\begin{center}
  \includegraphics[width=15cm]{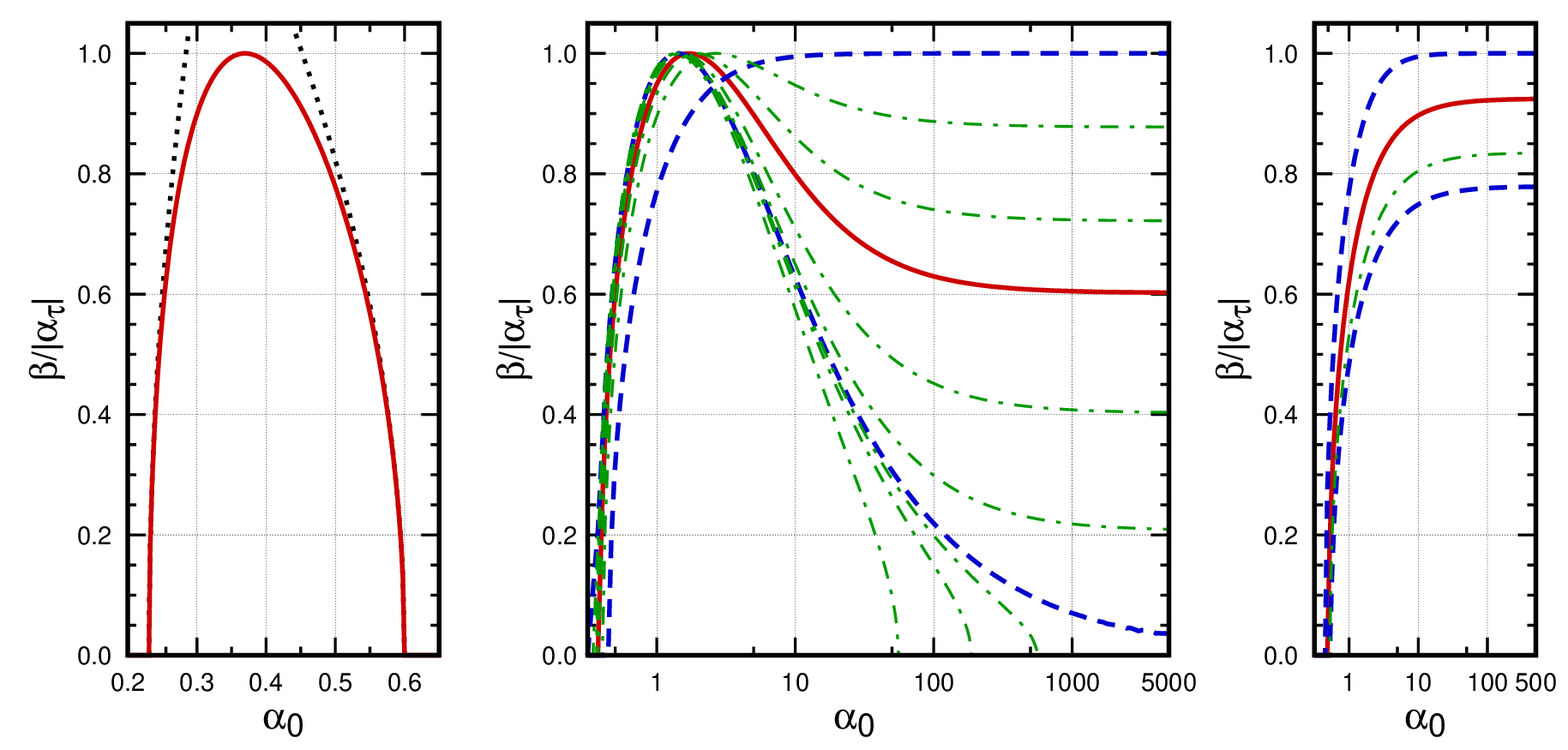}
\end{center}
\caption{The relative value $\beta/|\alpha_\tau|$ for a~hexagon with FM~bonds given
by~\eref{hexadata} as a~function of~$\alpha_0$; in the left panel the asymptotes
determined in Eqs.\ \eref{BetaAsy} and \eref{betaAsy1} are also plotted with dotted curves.
Red solid curves correspond to the values of $\alpha_\tau$ used in the previous figures, i.~e.,
$|\alpha_\tau|=1/3$ (the left panel, type~II), $|\alpha_\tau|=4/5$ (the middle panel, type
I$_\beta$), and $|\alpha_\tau|=3/2$ (the right panel, type I$_\alpha$). Blue dashed curves
show behaviour of $\beta/|\alpha_\tau|$ for $|\alpha_\tau|=\alpha_{\rm(d)}$ (the middle panel),
$|\alpha_\tau|=\alpha_{\rm(d')}$ (the middle and right panels), and
$|\alpha_\tau|=\alpha_\sigma$ (the right panel) (see theorem~\ref{TPD}). Green
dash-dotted curves are plotted for intermediate values of~$|\alpha_\tau|$.
A log-linear scale is used in the middle and  right panels.
\label{n6beta}}
\end{figure}

This system is discussed for three different values of
$|\alpha_\tau|=|\alpha_1|=1/3,4/5, 3/2$ corresponding to three types of the
$\alpha_0$-family (II, I$_\beta$, I$_\alpha$, respectively) defined in
theorem~\ref{TPD}. Basic parameters of the system considered for these three
cases are gathered in \tref{HexParam} whereas some dependencies are depicted in
the subsequent figures \ref{n6ener}--\ref{n6beta}.

The behaviour of the ground state energy $E$ as a~function of~$\alpha_0$ is different
in types II and~I, see \fref{n6ener} (for types I$_\alpha$ and I$_\beta$ the graphs are analogous).
In accordance with \eref{EnDiff} energies of the collinear LECs are equal for
$\alpha_0=|\alpha_\tau|$. For type~II, i,~e., for $\alpha_0\ge \alpha_{\rm{(c')}}$ the LEC
energy is given by \eref{EPsiUpDown}. This line represents always an upper bound of $E\left(\alpha_0\right)$;
only for the special value of $\left|\alpha_1\right|=\alpha_{\rm (d)}=3/4$ it is an asymptote to the energy graph.

In \fref{n6angl} the dependencies of angles $\Psi=-\psi_0$, $\psi_1$, and
$\psi_2$ on $\alpha_0$ are presented. Behaviour of other angles is analogous to
the $\psi_2$~graph, cf.~\eref{defBeta}. Note the maxima of $\psi_2$ in type~II
(left panel) and type~I$_\beta$ (middle panel). The angle~$\psi_1$ reaches
the value~$\pi$ in type~II only, in accordance with \eref{newpsi0coplanar}.

The last \fref{n6beta} shows changes in the dependence of the Lagrange parameter~$\beta$
on the antiferromagnetic coupling~$\alpha_0$. This parameter equals zero in collinear
LECs, i.e., for $\alpha_0\le\alpha_{\rm (c)}$ and, in the type~II only, for
$\alpha_0\ge\alpha_{\rm (c')}$; see the left panel and three lower dash-dotted
curves in the middle panel. It reaches its maximum in the cases of
type~II and  type~I$_\beta$, i.~e., for $0<|\alpha_\tau|<\alpha_{\rm (d')}$,
see the left and middle panel in \fref{n6beta}. After that  $\beta$
decreases monotonically  until it reaches the value $0$ (type II)
or approaches its limit value $\beta_\infty$ (type~I$_\beta$).
In the case of type~I$_\alpha$ the function $\beta(\alpha_0)$ has no maximum
but is always monotonically increasing and approaches its limit value
$\beta_\infty$, see the right panel in  \fref{n6beta}.
The values of $\beta_\infty/|\alpha_\tau|$ that can be observed in \fref{n6beta}
are in accordance with those given in \tref{HexParam} and with the equations
\eref{PsiprimeLimit} and \eref{betaLimit} given in proposition \ref{PropLimBeta}.

\section{Summary}\label{Sum}
The problem of finding all ground states of the $N$-spin polygon depends on an
initially confusing set of $N$~parameters, the exchange integrals
$\alpha_0,\ldots,\alpha_{N-1}$. A~certain simplification is achieved by
a~repeated spin flip operation, which reduces the problem to the special case
$\alpha_0>0$, $\alpha_1<0,\ldots,\alpha_{N-1}<0$. An important intermediate
result states that all ground states are either collinear or coplanar. This
allows the ground states to be parameterized by $N$~difference angles, which can
be written as functions of a~single Lagrange parameter~$\beta$.

Our investigations suggest that the AFM coupling $\alpha_0$ and the weakest FM
coupling $\alpha_\tau$ should be considered as variables, while the remaining FM
bonds are regarded as fixed. The resulting two-dimensional phase diagrams are all
qualitatively similar with the structure shown as an example in \fref{FIGPD}.

First, there are two separate phases with the collinear LEC
$\bPsi_{\uparrow\ldots\uparrow}$ and $\bPsi_{\uparrow\ldots\downarrow}$
which prevail for small $\alpha_0$ and $\left|\alpha_\tau\right|$, respectively.
There is a~large region of coplanar LEC in between, so that the phase boundaries
are symmetrical with respect to the reflection at the diagonal
$|\alpha_\tau|=\alpha_0$ and can be analytically calculated. They approach the
straight line $\alpha_0=\alpha_{\rm (d)}$ ($|\alpha_\tau|=\alpha_{\rm (d)}$,
respectively), asymptotically, where the characteristic value $\alpha_{\rm (d)}$
can be calculated in terms of the remaining FM bonds. It has to be emphasized
that competing interactions are present in the system considered for all
$\alpha_0>0$. Nevertheless, collinear LECs, characteristic for non-frustrated
systems, are observed. In both cases, i.e., for $\bPsi_{\uparrow\ldots\uparrow}$
and $\bPsi_{\uparrow\ldots\downarrow}$, a~single bond is not satsified, whereas
in the coplanar LEC we have $E_\mu>-|\alpha_\mu|$ for all $0\le \mu<N$, cf.\
\eref{PolCI}, and hence no bond is satisfied.

Other significant quantities are the energy $E(\alpha_0)$, the Lagrange
parameter~$\beta(\alpha_0)$ and the difference angle $\Psi(\alpha_0)$
between the spins at sites $\mu=0$ and $\mu=1$, each considered as a~function
of~$\alpha_0$. The energy~$E(\alpha_0)$ is linear in the collinear phases and
has a maximum as its only extreme value. With $\beta(\alpha_0)$, the maximum
only results for the case $0<|\alpha_\tau|<\alpha_{\rm (d')}$, where the second
characteristic value $\alpha_{\rm (d')}$ is again a~function of the remaining
FM~bonds. These remarks complete the general description of the LECs for spin polygons
depending on the $N$-exchange integrals $\alpha_0,\ldots,\alpha_{N-1}$.

Let us add that our theory can be directly applied to certain other spin systems
composed of polygons and spin chains. These systems can be characterized as ``trees"
in the graph-theoretic sense, whose vertices are given by polygons
(of possibly different sizes including $N=1$)
connected by edges that are given by finite spin chains.
These spin chains can also have zero length, in which case two polygons are
glued together in such a way that they have exactly one common spin site.
Due to their rotational degeneracy, the local ground states in the polygons
of such systems can always be extended to a global ground state.


\appendix
\section{Proofs}
\subsection{Proposition \ref{Pknull}(i)}
Recall that
\begin{equation}\label{Energy}
 E(\bPsi)=\alpha_0\,\cos\Psi-\sum_{\mu=1}^{N-1}|\alpha_\mu|\,\cos\psi_\mu
 \;,
\end{equation}
see \eref{EnbyPsi},
and that $\psi_0\neq 0$ for any coplanar LEC,
see \eref{newpsi0coplanar}.
First we consider the
\begin{itemize}
  \item Case $-\pi<\psi_0<0$:
\end{itemize}
Since $\psi_0> -\pi$ and $\psi_\mu\ge 0$ for all $0<\mu<N$, see \eref{psimuge0},
we have, according to \eref{AngCons},
\begin{equation}\label{knotnegative}
-\pi < \sum_{\mu=0}^{N-1}\psi_\mu = \Psi +\psi_0=2 k \pi
\;,
\end{equation}
which means that $k\in {\mathbb Z}$ can only take the values $k=0,1,2,\ldots$.
In order to derive a~contradiction we assume that
\begin{equation}\label{PsiContra}
  \Psi =\sum_{\mu=1}^{N-1}\psi_\mu=2\,k\,\pi-\psi_0=2\,k\,\pi+\left|\psi_0\right|,
    \quad {\rm for}\quad k\in{\mathbb N} \quad{\rm and }\quad k\ge 1.
\end{equation}
Upon defining
\begin{equation}\label{defgamma}
  \gamma:= \frac{\left|\psi_0\right|}{2\,k\,\pi+\left|\psi_0\right|},
    \quad {\rm such \,\,that}\quad  0<\gamma<1,
\end{equation}
we obtain
\begin{equation}\label{PsiTilde}
 \widetilde{\Psi}:= \sum_{\mu=1}^{N-1}\left(\gamma\,\psi_\mu \right)
 =\gamma\,\sum_{\mu=1}^{N-1}\psi_\mu
 \stackrel{(\ref{PsiContra},\,\ref{defgamma})}{=}\left|\psi_0\right|
\;.
\end{equation}
The new spin configuration
$\widetilde{\bPsi}=(\gamma\psi_1,\ldots,\gamma\psi_{N-1})$ has the energy
\begin{equation}\label{newenergy}
\fl  E\left(\widetilde{\bPsi}\right) = \alpha_0\cos\widetilde{\Psi}
    -\sum_{\mu=1}^{N-1}
    |\alpha_\mu|\cos(\gamma \psi_\mu)
  \stackrel{\eref{PsiTilde}}{=}\alpha_0\cos \Psi
    -\sum_{\mu=1}^{N-1}|\alpha_\mu|
    \cos(\gamma \psi_\mu)< E(\bPsi),
\end{equation}
using in the last inequality that $\cos(\gamma \psi_\mu)>\cos\psi_\mu$ for all $0<\mu<N$
due to $0< \psi_\mu \le\pi$, see \eref{psimuge0} and \eref{newpsi0coplanar}.
This contradicts
the assumption that $\bPsi$ is an LEC. We conclude that $k\ge 1$ in
\eref{PsiContra} cannot hold, so $k=0$. With this information we rewrite \eref{PsiContra} as
\begin{equation}\label{knullx}
  \Psi =\sum_{\mu=1}^{N-1}\psi_\mu=\left| \psi_0\right| =-\psi_0,
\end{equation}
which proves \eref{knull}. Finally, \eref{Psirange} follows from \eref{knullx},
and $-\pi < \psi_0 <0$.\\

Next we consider the
\begin{itemize}
  \item Case $0<\psi_0<\pi$:
\end{itemize}
This case will not occur but cannot be excluded from the outset.
In order to derive a~contradiction we assume that
\begin{equation}\label{PsiContra2}
  \Psi =\sum_{\mu=1}^{N-1}\psi_\mu=2\,k\,\pi-\psi_0,
    \quad {\rm for}\quad k\in{\mathbb N} \quad{\rm and }\quad k\ge 1
    \;,
\end{equation}
and define
\begin{equation}\label{defgamma2}
  \gamma:= \frac{\psi_0}{2\,k\,\pi-\psi_0}
  \;.
\end{equation}
Due to $\psi_0 < \pi \le 2k\pi$ the denominator of \eref{defgamma2} is positive
and hence $\gamma>0$. Moreover, the implications
\begin{equation}\label{estimatepsi0}
 \psi_0 < \pi \le k\pi \quad\Rightarrow \quad 2\psi_0 < 2 k \pi \quad\Rightarrow \quad \psi_0 < 2k\pi -\psi_0
\end{equation}
show that $\gamma<1$. For the next part of the proof we can analogously
proceed as for the case $-\pi<\psi_0<0$ in order to conclude
\begin{equation}\label{knullx2}
  \Psi =\sum_{\mu=1}^{N-1}\psi_\mu=-\psi_0
  \;.
\end{equation}
But this implies $-\pi < \Psi <0$ and hence $\sin\Psi <0$ in contradiction
to \eref{Signsinpsi}, thereby completing the proof of Proposition  \ref{Pknull}(i).

\subsection{Proposition \ref{Pknull}(ii)}
In case of a~collinear LEC we can apply proposition~\ref{propCollLEC} and
conclude that, in the regular domain, all $\psi_\mu=0$ for $0<\mu<N$ such that
\eref{PsiRC} is satisfied. Therefore, we can turn to the case of coplanar LEC.

In order to derive a~contradiction, assume that for some $0<\lambda<N$ we have
\begin{equation}\label{LamdaContra}
  \psi_\lambda =\frac{\pi}{2}+\delta, \quad {\rm for\;\; some }\quad
    0<\delta\le \frac{\pi}{2}.
\end{equation}
We consider a~modified spin configuration $\widetilde{\bPsi}$ by setting
$\widetilde{\psi}_\lambda=\frac{\pi}{2}-\delta$ and leaving the other angles
unchanged. The energies of both configurations are given by
\begin{eqnarray}\label{EPsiDelta}
  E(\bPsi)&=\alpha_0\,\cos \Psi-\sum_{\mu=1,\,\mu\neq\lambda}^{N-1}
    \left|\alpha_\mu\right|\,\cos\psi_\mu
    +\left|\alpha_\lambda\right|\,\sin\delta,\\
\label{EPsiTildeDelta}
  E(\widetilde{\bPsi})&=\alpha_0\,\cos\left(\Psi-2\delta\right)
    -\sum_{\mu=1,\,\mu\neq\lambda}^{N-1}\left| \alpha_\mu\right|\,\cos\psi_\mu
    -\left| \alpha_\lambda\right|\,\sin\delta.
\end{eqnarray}
Their difference satisfies
\begin{eqnarray}
\nonumber
  E(\bPsi)-E(\widetilde{\bPsi})&=
    \alpha_0\left(\cos \Psi -\cos(\Psi-2\delta)\right)
      +2|\alpha_\lambda|\,\sin\delta\\
\label{Ediff2}
  &= 2\sin\delta \left(|\alpha_\lambda|-\alpha_0\,\sin(\Psi-\delta)\right)>0,
\end{eqnarray}
since
\[
  |\alpha_\lambda|\ge |\alpha_\tau|>\alpha_0>\alpha_0\,\sin(\Psi-\delta).
\]
In \eref{Ediff2} we have used the general trigonometric formula
\begin{equation}\label{trigono1}
 \cos \alpha - \cos \beta = -2 \sin\frac{\alpha+\beta}{2}\, \sin\frac{\alpha-\beta}{2}\
 \;,
\end{equation}
which, after substituting $\alpha=\Psi, \beta = \Psi-2\delta$ yields
\begin{equation}\label{trigono2}
 -2 \sin \frac{2\Psi-2\delta}{2}\, \sin\frac{2\delta}{2} = 2 \sin(\delta-\Psi)\, \sin\delta
 \;.
\end{equation}

Summarizing, $ E(\bPsi)>E(\widetilde{\bPsi})$ and hence $\bPsi$ cannot be an LEC,
contrary to the assumption. This completes the proof of the proposition \ref{Pknull}.

\subsection{Theorem \ref{ThAlphC}}
Let $\widehat{\bPsi}$ be an LEC. At $\bPsi=\widehat{\bPsi}$ the conditions
\eref{MinCond1} have to be satisfied and, additionally,
\begin{equation}\label{MinCond2}
  H(\widehat{\boldsymbol\Psi})\ge 0,
\end{equation}
where $H$~denotes the $(N-1)$-dimensional real, symmetric Hessian matrix of
$E(\bPsi)$, with entries
\begin{equation}\label{Hessian}
  H_{\mu\nu}
    =\frac{\partial^2 E({\boldsymbol\Psi})}{\partial\psi_\mu\partial\psi_\nu},\quad 0<\mu,\nu<N.
\end{equation}
The condition $H\ge 0$ means that all eigenvalues of~$H$ are non-negative, and
hence $\det H\ge 0$. By evaluating \eref{Hessian} we obtain
\begin{equation}\label{MinCond2eval}
  H_{\mu\nu}
    =|\alpha_\mu|\,\cos\psi_\mu\,\delta_{\mu\nu}-\alpha_0\,\cos\Psi,
\end{equation}
for all $0<\mu<N$. For later purposes, we write~$H$ in the form
\begin{equation}\label{Hessian2}
  H=A+\bi{u}\,\bi{v}^{\sf T},
\end{equation}
where $A$ is the diagonal matrix with entries
$A_{\mu\mu}=|\alpha_\mu|\,\cos \psi_\mu$, $\bi{u}$ a~column vector
with constant entries $\sqrt{\alpha_0\,|\cos\Psi|}$ and
$\bi{v}=-\sgn(\cos\Psi)\,\bi{u}$.
\medskip

\noindent 1\@. \textsl{Only-if-part}\\
We assume that the LEC $\widehat{\bPsi}$ is collinear, i.e.,
\begin{equation}\label{LECcoll}
  \sin \widehat{\Psi}=\sin\widehat{\psi}_\mu=0
\end{equation}
for all $0<\mu<N$. By proposition~\ref{propCollLEC} we conclude
\begin{equation}\label{LECcoll2}
  \widehat{\psi}_\mu = \widehat{\Psi}=\sum_{\nu=1}^{N-1} \widehat{\psi}_\nu=0,
\end{equation}
for all $0<\mu<N$. It follows that the diagonal matrix~$A$ in \eref{Hessian2}
satisfies $A_{\mu\mu}=\left|\alpha_\mu\right|>0$ for all $0<\mu<N$ and is
hence invertible. Therefore, by the matrix determinant lemma~\cite{MatAlg},
\begin{equation}\label{matdetlemma1}
  \det H= \left(1+\bi{u}^{\sf T}\,A^{-1}\,\bi{v}\right)\,\det A,
\end{equation}
which at $\bPsi=\widehat{\bPsi}$ reduces to
\begin{equation}\label{matdetlemma2}
  \det H\left(\widehat{\bPsi}\right)= \left(1-\alpha_0\,\sum_{\mu=1}^{N-1}
    |\alpha_\mu|^{-1}\right)\,\prod_{\mu=1}^{N-1}|\alpha_\mu|,
\end{equation}
using \eref{LECcoll2}. The condition $\det H(\widehat{\bPsi})\ge 0$, see
\eref{MinCond2}, then implies
\begin{equation}\label{claim}
  0<\alpha_0 \le \left(\sum_{\mu=1}^{N-1}|\alpha_\mu|^{-1}\right)^{-1}
    =\alpha_{\rm (c)}.
\end{equation}
Note, further, that $|\alpha_\mu|<|\alpha_\tau|$ for all $0<\mu<N$ implies
$\sum_{\mu=1}^{N-1}|\alpha_\mu|^{-1}>1/|\alpha_\tau|$ and hence
$\alpha_{\rm (c)}<|\alpha_\tau|$, which concludes the only-if-part of the proof.
\medskip

\noindent 2\@. \textsl{If-part}\\
We have to prove that any LEC is collinear for
$0<\alpha_0 \le \alpha_{\rm (c)}$. It is straightforward to show that the
collinear state $\bPsi_{\uparrow\ldots\uparrow}$ given by $\psi_\mu=0$ for
all $0<\mu<N$ satisfies \eref{MinCond1} and \eref{MinCond2}, and is hence
a~plausible candidate for an LEC (note that these conditions are only
necessary). It remains to show that there is no coplanar state also
satisfying \eref{MinCond1} and~\eref{MinCond2}.

In order to derive a~contradiction we assume that $\bPsi$ is such a~state.
Due to \eref{Psirange} it satisfies
\begin{equation}\label{Psipos}
  0< \Psi < \pi.
\end{equation}
By virtue of \eref{MinCond1} and \eref{MinCond1eval} we have
\begin{equation}\label{sinpsi}
  \sin\psi_\mu=\frac{\alpha_0}{|\alpha_\mu|}\,\sin\Psi,
\end{equation}
for all $0<\mu<N$. By means of proposition~\ref{Pknull} it holds that
\begin{equation}\label{psirestrict}
  0< \psi_\mu \le \pi/2,\qquad {\rm for\;\; all} \quad 0<\mu<N,
\end{equation}
where the first strict inequality is due to the LEC being coplanar, see \eref{newpsi0coplanar}.
In this case
the function $\arcsin$ is well-defined for all arguments~$\psi_\mu$ and we may
write
\begin{equation}\label{deff}
  \Psi = \sum_{\mu=1}^{N-1}\psi_\mu \stackrel{\eref{sinpsi}}{=}
    \sum_{\mu=1}^{N-1}\arcsin\left(\frac{\alpha_0}{|\alpha_\mu|}\,
    \sin\Psi\right)  =: f(\Psi),
\end{equation}
where~$f$ is a~smooth function $f\colon [0,\pi] \to \mathbb{R}$ satisfying
$f(0)=0$. After some calculations we obtain
\begin{equation}\label{fder1}
  \frac{\rmd f}{\rmd \Psi}
    =\sum_{\mu=1}^{N-1}\frac{\alpha_0}{|\alpha_\mu|}\cos\Psi
     \left(1-\left(\frac{\alpha_0}{|\alpha_\mu|}
     \sin\Psi\right)^2\right)^{-1/2},
\end{equation}
and
\begin{equation}\label{fder2}
   \frac{\rmd^2 f}{\rmd\Psi^2}=\sum_{\mu=1}^{N-1}
\frac{\alpha_0\,\sin\Psi}{\left(\alpha_\mu^2-\alpha_0^2\sin^2\Psi\right)^{3/2}}\,
  \left(\alpha_0^2-\alpha_\mu^2\right).
\end{equation}
It follows that
\begin{equation}\label{fder10}
   \left.\frac{\rmd f}{\rmd \Psi}\right|_{\Psi=0}
    =\sum_{\mu=1}^{N-1 }\frac{\alpha_0}{|\alpha_\mu|}
    =\frac{\alpha_0}{\alpha_{\rm (c)}}\le 1.
\end{equation}
Moreover,
\begin{equation}\label{fconcave}
  \frac{\rmd^2 f}{\rmd\Psi^2}<0 \quad \mbox{for all}\quad 0< \Psi<\pi,
\end{equation}
due to the terms $\left(\alpha_0^2-\alpha_\mu^2 \right)<0$ in the sum
\eref{fder2}. This means that $f$~is a~strictly concave function passing
through the origin and having there a~slope less than or equal to~1. This
implies $f(\Psi)<\Psi$ for all $0<\Psi<\pi$ and hence \eref{deff} cannot be
satisfied. This completes the proof of the if-part and thereby of the theorem \ref{MTh}.

\subsection{Proposition \ref{Propunique}(i)}\label{proofPropunique1}
Define $F\colon [0,\pi] \to {\mathbb R}$ by $F(\Psi):=f(\Psi)-\Psi$. Hence
a~coplanar LEC $\widehat{\bPsi}$ yields a~zero of $F$ by means of
$F(\widehat{\Psi})=f(\widehat{\Psi})-\widehat{\Psi}=0$. Obviously, $F$~is also
strictly concave since
\begin{equation}\label{Fconcave}
  \frac{\rmd^2 F}{\rmd \Psi^2}<0
\end{equation}
for all $0<\Psi<\pi$, and satisfies
\begin{equation}\label{Fder0}
  \left.\frac{\rmd F}{\rmd\Psi}\right|_{\Psi=0}>0,
\end{equation}
by means of \eref{fder11}. We claim that
\begin{equation}\label{dFless0}
  \left.\frac{\rmd F}{\rmd\Psi}\right|_{\Psi=\widehat{\Psi}}<0.
\end{equation}
Assume the contrary
$\left.\frac{\rmd F}{\rmd\Psi}\right|_{\Psi=\widehat{\Psi}}\ge 0$ then
we have
\begin{equation}\label{Finta}
 \left.\frac{\rmd F}{\rmd\Psi}\right|_{\Psi=\widehat{\Psi}}-  \frac{\rmd F}{\rmd\Psi}
  = \int_{\Psi}^{\widehat{\Psi}} \underbrace{\left.\frac{\rmd^2 F}{\rmd \Psi^2}\right|_{\Psi=\Psi'}}
 _{\stackrel{\eref{Fconcave}}{<}0} d\Psi' <0
 \;.
\end{equation}
and hence
\begin{equation}\label{Fge0}
 \frac{\rmd F}{\rmd\Psi}>\left.\frac{\rmd F}{\rmd\Psi}\right|_{\Psi=\widehat{\Psi}} \ge 0
 \;.
\end{equation}
This implies
\begin{equation}\label{Fint1}
  F(\widehat{\Psi})=\int_{0}^{\widehat{\Psi}}
    \frac{\rmd F}{\rmd\Psi}\,\rmd \Psi >0,
\end{equation}
which contradicts $F(\widehat{\Psi})=0$, thereby proving \eref{dFless0}.

Now assume $F(\widehat{\Psi}_1)=F(\widehat{\Psi}_2)=0$ for some
$\widehat{\Psi}_1<\widehat{\Psi}_2$. From
$\left.\frac{\rmd F}{\rmd \Psi}\right|_{\Psi=\widehat{\Psi}_1}<0$,
see \eref{dFless0}, and \eref{Fconcave} we conclude
$\left.\frac{\rmd F}{\rmd\Psi}\right|_{\Psi=\Psi_0}<0$ for all
$\widehat{\Psi}_1<\Psi_0<\widehat{\Psi}_2$. Hence
\begin{equation}\label{Fint2}
  F(\widehat{\Psi}_2)=\underbrace{F(\widehat{\Psi}_1)}_{=0}+
    \int_{\widehat{\Psi}_1}^{\widehat{\Psi}_2}
    \frac{\rmd F}{\rmd\Psi}\,\rmd\Psi <0,
\end{equation}
which contradicts $F(\widehat{\Psi}_2)=0$. Hence any non-trivial zero of~$F$ is
unique.

\subsection{Proposition \ref{Propunique}(ii)}\label{proofPropunique2}
We reconsider the analytic function~$F$ defined in \ref{proofPropunique1} and
rewrite it in the form
$F\colon \left(\alpha_{\rm (c)},|\alpha_\tau|\right)\times(0,\pi)\to{\mathbb R}$
with
\begin{equation}\label{defF2}
  F\left(\alpha_0,\Psi\right) := \sum_{\mu=1}^{N-1}\arcsin
   \left(\frac{\alpha_0}{|\alpha_\mu|}\,\sin\Psi\right)-\Psi.
\end{equation}
Let $\widehat{\bPsi}$ be a~coplanar LEC such that
$\widehat{\Psi}=\Psi(\alpha_0)$. To apply the inverse function theorem (IFT)
we have to show that $\partial F/\partial \Psi\neq 0$ at $\Psi=\widehat{\Psi}$.
In fact this follows from
\begin{equation}\label{partialFPsi}
  \left.\frac{\partial F}{\partial\Psi}\right|_{\Psi=\widehat{\Psi}}<0,
\end{equation}
see \eref{dFless0}. The analytic IFT then yields the existence of a~local
analytical function
$\psi\colon(\alpha_0-\delta,\alpha_0+\delta)\to {\mathbb R}$ for some $\delta>0$
satisfying $F\left(\alpha,\psi(\alpha)\right)=0$, for all
$\alpha_0-\delta < \alpha< \alpha_0+\delta$, see \cite{AnalIFC}. According to
proposition~\ref{Propunique} the function~$\psi$ locally coincides with~$\Psi$,
i.e., $\psi(\alpha)=\Psi(\alpha)$ for all
$\alpha_0-\delta<\alpha<\alpha_0+\delta$.

\subsection{Proposition \ref{Propunique}(iii)}\label{proofPropunique3}

The proof for his case is completely analogous to the preceding one
if we can show that the given bounds for $\left| \alpha_{N-\tau}\right|$ imply
that we are in the regular domain $\alpha_{\rm (c)}< \alpha_0 < \left| \alpha_\tau\right|$.
The second inequality  $\alpha_0 < \left| \alpha_\tau\right|$ is satisfied by assumption.
To prove the first inequality $\alpha_{\rm (c)}< \alpha_0$ we make a case distinction.
\paragraph{Case of $\frac{1}{\alpha_0}>{\sum_\mu^\ast}\frac{1}{\left|\alpha_\mu\right|}$}:
The first inequality follows from the equivalences
\begin{eqnarray}
\nonumber
   && \left|\alpha_{N-\tau}\right|\stackrel{\eref{conditionAlphaNminusTau}}{<}
   \left(\frac{1}{\alpha_0}-{{\sum_\mu}^\ast}\frac{1}{\left|\alpha_\mu\right|}\right)^{-1} \\
   \nonumber
  &\Leftrightarrow&\frac{1}{\left|\alpha_{N-\tau}\right|}>\frac{1}{\alpha_0}-
  {{\sum_\mu}^\ast}\frac{1}{\left|\alpha_\mu\right|}\\
   \nonumber
  &\Leftrightarrow& \frac{1}{\alpha_{\rm (c)}}=
  \sum_{\mu=1}^{N-1}\frac{1}{\left|\alpha_\mu\right|} > \frac{1}{\alpha_0}\\
  \label{equiv1}
  &\Leftrightarrow& \alpha_{\rm (c)} < \alpha_0
  \;.
\end{eqnarray}
\paragraph{Case of $\frac{1}{\alpha_0}\le{\sum_\mu^\ast}\frac{1}{\left|\alpha_\mu\right|}$}:
The first inequality follows from
\begin{eqnarray}
\nonumber
   && \frac{1}{\alpha_0}\le{{\sum_\mu}^\ast}\frac{1}{\left|\alpha_\mu\right|}
   = \sum_{\mu=1}^{N-1}\frac{1}{\left|\alpha_\mu\right|} -
   \frac{1}{\left|\alpha_{N-\tau} \right|}=\frac{1}{\alpha_{\rm (c)}}-
   \frac{1}{\left|\alpha_{N-\tau} \right|}< \frac{1}{\alpha_{\rm (c)}}\\
   \label{equiv2}
  &\Rightarrow&\alpha_{\rm(c)} < \alpha_0
  \;.
\end{eqnarray}
In this case, we therefore do not need to set a finite upper limit
for $\left|\alpha_{N-\tau}\right|$ to ensure that we are in the regular domain.
\subsection{Theorem \ref{TPD}}\label{proofTPD}
\paragraph{Part (i)}
The function
  \begin{equation}\label{defG}
    g(x):= \left.\sum\right.'_{\mu} \arcsin \frac{x}{|\alpha_\mu|}
  \end{equation}
  is continuous and monotonically strictly increasing for
  $0\le x\le|\alpha_\sigma|$ and hence invertible. Moreover,
  $g\left(|\alpha_\sigma|\right)>\pi/2$ since the term with $\mu=\sigma$ in the
  sum~\eref{defG} yields~$\pi/2$ and, for $N>3$, the remaining terms are
  positive. Thus there exists a unique $0<x_1<|\alpha_\sigma|$ such that
  $g(x_1)=\pi/2$. It follows that $\alpha_{\rm (d')}=x_1$ and
  $0<\alpha_{\rm (d')}<|\alpha_\sigma|$. To show
  $\alpha_{\rm(d)}<\alpha_{\rm(d')}$ insert $x=\alpha_{\rm(d)}<|\alpha_\sigma|$
  into $g(x)$ and use that
  \begin{equation}\label{boundArcsin}
     0<\arcsin y < y\,\frac{\pi}{2}
  \end{equation}
  for all $0<y<1$. This entails
  \begin{eqnarray}\label{inequality1}
   g\left(\alpha_{\rm(d)} \right)&\stackrel{\eref{defG}}{=}
   \left.\sum\right.'_{\mu} \arcsin
   \frac{\alpha_{\rm(d)}}{|\alpha_\mu|}\stackrel{\eref{boundArcsin}}{<}
   \left.\sum\right.'_{\mu}
   \frac{\alpha_{\rm(d)}}{|\alpha_\mu|}\,\frac{\pi}{2}\\
   \label{inequality2}
   &=
   \frac{\pi}{2}\,\alpha_{\rm(d)} \,\left.\sum\right.'_{\mu}
   \frac{1}{|\alpha_\mu|}\stackrel{\eref{Alphf}}{=}\frac{\pi}{2}=
   g\left(\alpha_{\rm(d')} \right),
  \end{eqnarray}
  whence $\alpha_{\rm(d)}<\alpha_{\rm (d')}$ since $g$ is strictly monotonically
  increasing.
\paragraph{Part (ii)}
    As noted above we have the equivalent statements
  \begin{eqnarray}
  \nonumber
    & {\rm The}\;\;\alpha_0\mbox{-}{\rm family\; is\; of\;type\; II} \\
     \label{equivalent1}
   \Leftrightarrow\quad & 0< \alpha_{\rm (c')}\stackrel{\eref{retrans7a}}{=}
   \left(|\alpha_\tau|^{-1}-\alpha_{\rm (d)}^{-1}\right)^{-1}<\infty \\
      \label{equivalent2}
   \Leftrightarrow& |\alpha_\tau| < \alpha_{\rm (d)}.
  \end{eqnarray}
\paragraph{Part (iii)}
We have the equivalent statements
  \begin{eqnarray}
  \nonumber
    & {\rm The}\;\alpha_0\mbox{-}{\rm family\; is\; of\;type\; I}_\beta \\
    \label{equivalent3}
 \Leftrightarrow\quad& 0< \alpha_{\rm (bm))}\stackrel{\eref{retrans5}}{=}
    |\alpha_\tau|\left(\cos \left.\sum\right.'_{\mu}
    \arcsin\frac{|\alpha_\tau|}{|\alpha_\mu|}\right)^{-1}
    <\infty \\
       \label{equivalent4}
 \Leftrightarrow& g\left(|\alpha_\tau|\right)\stackrel{\eref{defG}}{=}
     \left.\sum\right.'_{\mu}\arcsin
     \frac{|\alpha_\tau|}{|\alpha_\mu|}<\frac{\pi}{2}
  \stackrel{\eref{defAlphaG}}{=}g\left(\alpha_{\rm (d')}\right)\\
      \label{equivalent5}
 \Leftrightarrow& |\alpha_\tau| < \alpha_{\rm (d')}.
  \end{eqnarray}
  In the limit $|\alpha_\tau|\to\alpha_{\rm (d')}$ we have
  $\alpha_{\rm (bm)}\to\infty$ and hence the case
  $|\alpha_\tau|=\alpha_{\rm (d')}$ belongs to the type I$_\alpha$.

\paragraph{Part (iv)}
The remaining statement follows since (ii)--(iv) is a~complete case distinction.

\subsection{Proposition \ref{PropLimBeta}}\label{proofLimBeta}
We consider \eref{deff} for the primed quantities and conclude
\begin{eqnarray}
 \label{eqPsiprime1}
  \Psi' &= \sum_{\mu=1}^{N-1}\arcsin\left(
  \frac{\left|\alpha_0'\right|}{\left|\alpha_\mu'\right|}\sin\Psi' \right)\\
  \label{eqPsiprime2}
   &= \sum_{\mu=1,\,\mu\neq N-\tau}^{N}\arcsin\left(
   \frac{\left|\alpha_0'\right|}{\left|\alpha_\mu'\right|}\sin \Psi'\right)
   +\underbrace{\arcsin\left(\frac{\left|\alpha_0' \right|}
   {\left|\alpha_{N-\tau}' \right|} \sin \Psi'\right)}_{\to 0},
\end{eqnarray}
for $\left|\alpha_{N-\tau}' \right|=|\alpha_0|\to \infty$. Hence $\Psi_\infty'$
is a~solution of the equation
\begin{eqnarray}
\nonumber
  \Psi_\infty' &=  \sum_{\mu=1,\,\mu\neq N-\tau}^{N}\arcsin\left(
  \frac{\left|\alpha_0'\right|}{\left|\alpha_\mu'\right|}
    \sin \Psi_\infty'  \right) \\
\nonumber
   &= \sum_{\mu=0,\,\mu\neq N-\tau}^{N}\arcsin\left(
  \frac{\left|\alpha_0' \right|}{\left|\alpha_\mu' \right|}\sin \Psi_\infty'
  \right)- \arcsin \left( \sin \Psi_\infty'\right) \\
  \label{eqPsiprime5}
  &=\sum_{\mu=1}^{N}\arcsin\left(
  \frac{|\alpha_\tau|}{|\alpha_\mu|}\sin \Psi_\infty'
  \right)- \arcsin \left( \sin \Psi_\infty'\right).
\end{eqnarray}

By applying proposition \ref{Propunique}(iii) to the primed quantities
we conclude that $\Psi'$ will be an analytic function of $|\alpha_{N-\tau}'|=\alpha_0$
for some open interval with upper limit either
$b:=\left(\frac{1}{\alpha_0'}-\sum_{\mu}^\ast \frac{1}{\left|\alpha_\mu' \right|}\right)^{-1}$
or  $\infty$. Let us evaluate $b$ in terms of the unprimed quantities:
\begin{eqnarray}
\nonumber
  b &=&\left(\frac{1}{\alpha_0'}-\sum_{\mu}^\ast \frac{1}{\left|\alpha_\mu' \right|}\right)^{-1} \\
  \nonumber
   &=& \left(\frac{1}{\alpha_0'}-\sum_{\mu=0,\mu\neq N-\tau}^{N-1} \frac{1}{\left|\alpha_\mu' \right|}+\frac{1}{\alpha_0'}\right)^{-1} \\
   \nonumber
   &=& \left(\frac{2}{\left|\alpha_\tau\right|}-\sum_{\mu=1}^{N-1} \frac{1}{\left|\alpha_\mu \right|}\right)^{-1} \\
   \label{evalb1}
   &=& \left( \frac{2}{\left|\alpha_\tau\right|}- \frac{1}{\alpha_{\rm (c)}}\right)^{-1} \stackrel{\eref{retrans7}}{=}\alpha_{\rm(c')}
   \;.
\end{eqnarray}
It follows that the case distinction of \eref{conditionAlphaNminusTau} corresponds to the case distinction
between type I and type II $\alpha_0$-families. Since we are only dealing with type I in this proof
the upper limit of the interval where $\Psi'(\alpha_0)$ is analytical will be $\infty$.

Let us rewrite the r.~h.~s.~minus the l.~h.~s.~of \eref{eqPsiprime1} as
$ G\left(\alpha_0, \Psi' \right)=0$.
As in the proof of  proposition \ref{Propunique}(ii) we can show that
\begin{equation}\label{eqPsiprime8}
 \frac{\partial G\left(\alpha_0, \Psi' \right)}{\partial \Psi'}<0
 \;,
\end{equation}
and, further, that
\begin{equation}\label{eqPsiprime6}
 \frac{d \Psi'}{d\alpha_0} = -\frac{\partial G/\partial \alpha_0}{\partial G/\partial \Psi'}<0
 \;.
\end{equation}
The latter follows from
\begin{eqnarray}
\label{eqPsiprime7}
  \frac{\partial G}{\partial \alpha_0} &=& \frac{\partial}{\partial \alpha_0}
  \arcsin\left(\frac{\left|\alpha_0' \right|}{\left|\alpha_{N-\tau}' \right|}\sin \Psi' \right)
 = \frac{\partial}{\partial \alpha_0}
  \arcsin\left(\frac{\left|\alpha_\tau \right|}{\alpha_0}\sin \Psi' \right)
   \\
   &=& -\frac{\left|\alpha_\tau \right|\sin \Psi'}{\sqrt{\alpha_0^2-
   \left|\alpha_\tau \right|^2\,\sin^2 \Psi' }\,\alpha_0}<0
   \;,
\end{eqnarray}
and \eref{eqPsiprime5}.

Now we consider the $\alpha_0$-family of LEC. At $\alpha_0=  \left|\alpha_\tau \right|$
we have $\Psi'(\alpha_0)=\Psi(\alpha_0)>\pi/2$. For larger values $\alpha_0> \left|\alpha_\tau \right|$
the angle $\Psi'(\alpha_0)$ strictly decreases according to \eref{eqPsiprime6}
until it asymptotically approaches $\Psi'_\infty$ for $\alpha_0\to\infty$.
This means that for $\Psi_\infty'\ge \pi/2$ the maximum of $\beta(\alpha_0)$
which occurs at $\Psi'=\pi/2$  will not be attained. Conversely, for $\Psi_\infty'< \pi/2$
the maximum of $\beta(\alpha_0)$ will be attained. Thus we obtain the following
case distinction:

\paragraph{Type I$_\alpha$}
If the $\alpha_0$-family is of type I$_\alpha$ we have $\Psi_\infty'\ge \pi/2$
and hence $\arcsin\left( \sin \Psi_\infty'\right)= \pi-\Psi_\infty'$. Hence
\eref{eqPsiprime5} is equivalent to
\begin{equation}\label{{eqPsiprime6}}
 \pi = \sum_{\mu=1}^{N}\arcsin\left(
  \frac{|\alpha_\tau|}{|\alpha_\mu|}\sin \Psi_\infty' \right),
\end{equation}
whence \eref{PsiprimeLimitalpha} follows.

\paragraph{Type I$_\beta$}
If the $\alpha_0$-family is of type~I$_\beta$ we have $\Psi_\infty'<\pi/2$
and hence $\arcsin\left( \sin \Psi_\infty'\right)=\Psi_\infty'$. Hence
\eref{eqPsiprime5} is equivalent to
\begin{equation}\label{{eqPsiprime7}}
  \Psi_\infty' =\frac{1}{2} \sum_{\mu=1}^{N}\arcsin\left(
  \frac{|\alpha_\tau|}{|\alpha_\mu|}\sin \Psi_\infty'\right),
\end{equation}
whence  \eref{PsiprimeLimitbeta} follows.

\section{Limit point $\alpha_0=\left|\alpha_\tau\right|$}\label{Limit}
It is plausible that the results of \sref{RegDom} also hold for the limiting
case $\alpha_0\to|\alpha_\tau|$, especially that the LEC is coplanar in this
case. In fact, a~careful re-inspection of the proofs shows that they can be
extended to the case $\alpha_0=|\alpha_\tau|$ with one exception: The proof of
proposition~\ref{Pknull}(ii) is based on the fact that all collinear LECs in the
regular domain have $\psi_\mu=0$ for all $0\le\mu<N$. But for
$\alpha_0=|\alpha_\tau|$ there is collinear state
$\bPsi_{\uparrow\ldots\downarrow}$ which violates this condition, see
proposition~\ref{propCollLEC}. Although it is again plausible that
$\bPsi_{\uparrow\ldots\downarrow}$ cannot be an LEC for $\alpha_0=|\alpha_\tau|$
and hence proposition~\ref{Pknull}(ii) also holds in this case, it turns out to
be difficult to prove this without using the theory developed in \sref{RegDom}.
Therefore we have considered the limit point case in a~separate section.

Recall that at $\alpha_0=\alpha_{\rm (em)}$ the energy~$E(\alpha_0)$ has the
maximum. Since this is the only extremum of~$E(\alpha_0)$ in the coplanar phase
of the regular domain we conclude
\begin{equation}\label{Emax1}
  E(\alpha_0)\le E\left(\alpha_{\rm (em)}\right),\quad {\rm for\;all\;\;}
    \alpha_{\rm (c)}<\alpha_0<|\alpha_\tau|.
\end{equation}
For $\alpha_0=\alpha_{\rm (em)}$ the LEC will be coplanar. Hence
\begin{equation}\label{Emax2}
  E(\alpha_{\rm (em)})<E_{\rm max}^{\rm coll} :=
    E\left(\bPsi_{\uparrow\ldots\uparrow}\right)
    = E\left(\bPsi_{\uparrow\ldots\downarrow}\right),
\end{equation}
see proposition~\ref{propCollLEC}. Choose some~$\epsilon$ such that
\begin{equation}\label{Emax3}
  0<\epsilon<\case{1}{2}
    \left(E_{\rm max}^{\rm coll}-E(\alpha_{\rm (em)}\right).
\end{equation}
For the sake of clarity we will write the energy function with two arguments in
the form
\begin{equation}\label{EnergyFunction}
  \mathcal{E}(\alpha,\bPsi):=\alpha\,\cos
    \left(\sum_{\mu=1}^{N-1}\psi_\mu\right)
    -\sum_{\mu=1}^{N-1}|\alpha_\mu|\,\cos\psi_\mu.
\end{equation}

\begin{figure}[tb]
\begin{center}
  \includegraphics[width=10cm]{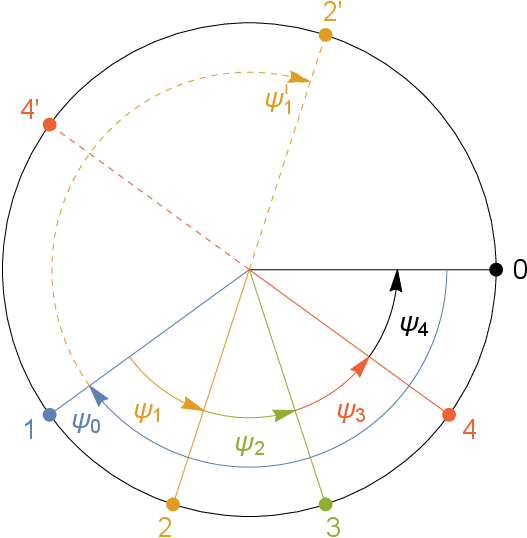}
\end{center}
\caption{
Illustration of our theory applied to the uniform AF pentagon.
Having performed the spin flips ${\mathbf s}_2\mapsto -{\mathbf s}_2$ and ${\mathbf s}_4\mapsto -{\mathbf s}_4$
the spin vectors of the resulting LEC configuration assume the positions marked by 0,1,2,3,4.
The difference angles satisfy $\sum_{\mu=0}^{4}\psi_\mu=0$ and $\psi_\mu=\pi/5$ for $\mu=1,2,3,4$,
hence $\psi_0=-4\pi/5$.
After reversing the spin flips (resulting in the new positions $2'$ and $4'$) the new difference angles
$\psi'_\mu,\,\mu=0,\ldots,4$ assume the constant value $-4\pi/5$.
For the sake of clarity, we have only shown the difference angle $\psi'_1$.
}
\label{penta}
\end{figure}

Next choose some $\alpha_0$ close to $|\alpha_\tau|$, i.e.,
\begin{equation}\label{Emax4}
  |\alpha_\tau|-\epsilon < \alpha_0 < |\alpha_\tau|,
\end{equation}
and take the corresponding LEC $\bPsi\left(\alpha_0\right)$ as a~trial state
for estimating the energy $E_{\rm min}(\left| \alpha_\tau\right|)$ of the LEC
at $|\alpha_\tau|$. We conclude
\begin{equation}\label{Emax5}
  \left|\mathcal{E}(\alpha_0,\bPsi_0)-\mathcal{E}
    (|\alpha_\tau|,\bPsi_0)\right|
    \stackrel{\eref{EnergyFunction}}{\le}
    \left|\left(\alpha_0-|\alpha_\tau|\right)\cos\Psi_0\right|
    \stackrel{\eref{Emax4}}{<}\epsilon.
\end{equation}
Further,
\begin{equation}\label{Emax6}
\fl E_{\rm min}\left(|\alpha_\tau|\right)
  \le{\mathcal E}\left(|\alpha_\tau|,\bPsi_0\right)
   \stackrel{\eref{Emax5}}{<}{\mathcal E}\left(\alpha_0,\bPsi_0\right)+\epsilon
   \stackrel{\eref{Emax1}}{\le}E\left(\alpha_{\rm (em)}\right)+\epsilon
   \stackrel{\eref{Emax3}}{<} E_{\rm max}^{\rm coll}.
\end{equation}

This estimate shows that the LEC at the limit point $|\alpha_\tau|$ cannot be
collinear and hence must be coplanar.\\

We will illustrate our theory using the well-known case of a pentagon with uniform AF coupling $\alpha=1$
which leads to a special case of  $\alpha_0=\left|\alpha_\tau\right|$. By the spin flips
${\mathbf s}_2\mapsto -{\mathbf s}_2$ and ${\mathbf s}_4\mapsto -{\mathbf s}_4$ this problem
is reduced to the case of $\alpha_0=1$ and $\alpha_1=\alpha_2=\alpha_3=\alpha_4=-1$.
The condition (\ref{defBeta}) implies
\begin{equation}\label{pentagon1}
\sin \Psi = \sin \psi_\mu,\quad \mbox{for all } \mu=1,...,4
\;.
\end{equation}
From $0\le \psi_\mu\le \pi/2$, see Prop.~\ref{Pknull} (ii), and (\ref{pentagon1}) we conclude that
all $\psi_\mu$ are equal, say, $\psi_1=\psi_2=\psi_3=\psi_4=:\psi$. It follows that
$-\psi_0=\Psi=\sum_{\mu=1}^4\psi_\mu=4\, \psi$, and (\ref{pentagon1}) implies
$\Psi=\pi-\psi=4 \psi$. Hence $\psi=\pi/5$ and $\Psi=4\pi/5$.
Reversing the spin flips
${\mathbf s}_2\mapsto -{\mathbf s}_2$ and ${\mathbf s}_4\mapsto -{\mathbf s}_4$
and denoting the resulting angles between the ground state vectors by $\psi'_\mu$
we obtain $\psi'_\mu=-4\pi/5$ for $\mu=0,\ldots,4$, which is in line with the well-known
result that the ground state of the AF pentagon is the spin pentagram, see Figure \ref{penta}.

\providecommand{\newblock}{}


\begin{thebibliography}{10}
\expandafter\ifx\csname url\endcsname\relax
  \def\url#1{{\tt #1}}\fi
\expandafter\ifx\csname urlprefix\endcsname\relax\def\urlprefix{URL }\fi
\providecommand{\eprint}[2][]{\url{#2}}

\bibitem{Bani18}
Baniodeh A, Magnani N, Lan Y, Buth G, Anson C~E, Richter J, Affronte M, Schnack
  J and Powell A~K 2018 {\em npj Quantum Materials\/} {\bf 3} 10

\bibitem{Maje18}
Majee M~C, Towsif~Abtab S~M, Mondal D, Maity M, Weselski M, Witwicki M,
  Bie{\'n}ko A, Antkowiak M, Kamieniarz G and Chaudhury M 2018 {\em Dalton
  Trans.\/} {\bf 47} 3425--39

\bibitem{Crai19}
Craig G~A, Velmurugan G, Wilson C, Valiente R, Rajaraman G and Murrie M 2019
  {\em Inorganic Chemistry\/} {\bf 58} 13815--25

\bibitem{Fu2020}
Fu Z, Qin L, Sun K, Hao L, Zheng Y~Z, Lohstroh W, G{\"u}nther G, Russina M, Liu
  Y, Xiao Y, Jin W and Chen D 2020 {\em npj Quantum Materials\/} {\bf 5} 32

\bibitem{Antk21}
Antkowiak M, Majee M~C, Maity M, Mondal D, Kaj M, Lesi{\'o}w M, Bie{\'n}ko A,
  Kronik L, Chaudhury M and Kamieniarz G 2021 {\em The Journal of Physical
  Chemistry\/} C {\bf 125} 11182--96

\bibitem{Shuk21}
Shukla P, Pal T~K, Sahoo S~C, Du M~H, Kong X~J and Das S 2021 {\em
  ChemistrySelect\/} {\bf 6} 2456--63

\bibitem{Tzio23}
Tziotzi T~G, Gracia D, Dalgarno S~J, Schnack J, Evangelisti M, Brechin E~K and
  Milios C~J 2023 {\em Journal of the American Chemical Society\/} {\bf 145}
  7743--7

\bibitem{Jin23}
Jin P~B, Luo Q~C, Gransbury G~K, Vitorica-Yrezabal I~J, Hajdu T, Strashnov I,
  McInnes E~J~L, Winpenny R~E~P, Chilton N~F, Mills D~P and Zheng Y~Z 2023 {\em
  Journal of the American Chemical Society\/} Published on-line 24 Nov. 2023

\bibitem{Dmit19}
Dmitriev D~V, Krivnov V~Y, Richter J and Schnack J 2019 {\em Phys. Rev.\/} B
  {\bf 99} 094410

\bibitem{Enge06}
Engelhardt L, Luban M and Schr{\"o}der C 2006 {\em Phys. Rev.\/} B {\bf 74}
  054413

\bibitem{Kons18}
Konstantinidis N~P 2018 {\em Journal of Magnetism and Magnetic Materials\/}
  {\bf 449} 55--62

\bibitem{Schu21}
Schubert D, Richter J, Jin F, Michielsen K, De~Raedt H and Steinigeweg R 2021
  {\em Phys. Rev.\/} B {\bf 104} 054415

\bibitem{Ueda17}
Ueda H, Okunishi K, Kr{\v{c}}m{\'a}r R, Gendiar A, Yunoki S and Nishino T 2017
  {\em Phys. Rev.\/} E {\bf 96} 062112

\bibitem{Gome18}
G{\'o}mez~Albarrac{\'{\i}}n F~A and Pujol P 2018 {\em Phys. Rev.\/} B {\bf 97}
  104419

\bibitem{Anis19}
Anisimov A~V, Pini M~G and Popov A~P 2019 {\em Journal of Magnetism and
  Magnetic Materials\/} {\bf 479} 105--13

\bibitem{GS23}
Gemb\'{e} M, Schmidt H~J, Hickey C, Richter J, Iqbal Y and Trebst S 2023 {\em
  Phys. Rev. Res.\/} {\bf 5} 043204

\bibitem{Furr13}
Furrer A and Waldmann O 2013 {\em Rev. Mod. Phys.\/} {\bf 85} 367--420

\bibitem{Timc13}
Timco G~A, McInnes E~J~L and Winpenny R~E~P 2013 {\em Chem. Soc. Rev.\/} {\bf
  42} 1796--806

\bibitem{Shar14}
Sharples J~W, Collison D, McInnes E~J~L, Schnack J, Palacios E and Evangelisti
  M 2014 {\em Nature Communications\/} {\bf 5} 5321

\bibitem{McIn15}
McInnes E~J~L, Timco G~A, Whitehead G~F~S and Winpenny R~E~P 2015 {\em Angew.
  Chem. Int. Ed.\/} {\bf 54} 14244--69

\bibitem{Ghir15}
Ghirri A, Chiesa A, Carretta S, Troiani F, van Tol J, Hill S, Vitorica-Yrezabal
  I, Timco G~A, Winpenny R~E~P and Affronte M 2015 {\em The Journal of Physical
  Chemistry Letters\/} {\bf 6} 5062--6

\bibitem{Bake16}
Baker M~L, Lancaster T, Chiesa A, Amoretti G, Baker P~J, Barker C, Blundell
  S~J, Carretta S, Collison D, G{\"u}del H~U, Guidi T, McInnes E~J~L,
  M{\"o}ller J~S, Mutka H, Ollivier J, Pratt F~L, Santini P, Tuna F,
  Tregenna-Piggott P~L~W, Vitorica-Yrezabal I~J, Timco G~A and Winpenny R~E~P
  2016 {\em Chemistry -- A European Journal\/} {\bf 22} 1779--88

\bibitem{Garl16}
Garlatti E, Bordignon S, Carretta S, Allodi G, Amoretti G, De~Renzi R,
  Lascialfari A, Furukawa Y, Timco G~A, Woolfson R, Winpenny R~E~P and Santini
  P 2016 {\em Phys. Rev.\/} B {\bf 93} 024424

\bibitem{Antk19}
Antkowiak M, Kamieniarz G, Timco G~A, Tuna F and Winpenny R~E~P 2019 {\em
  Journal of Magnetism and Magnetic Materials\/} {\bf 479} 166--9

\bibitem{Schn19}
Schnack J 2019 {\em Contemporary Physics\/} {\bf 60} 127--44

\bibitem{Maty21}
Matysiak J and Lema{\'n}ski R 2021 {\em Phys. Rev.\/} B {\bf 104} 014431

\bibitem{Poin23}
Pointillart F, Bernot K, Le~Guennic B and Cador O 2023 {\em Chem. Commun.\/}
  {\bf 59} 8520--8531

\bibitem{Chie24}
Chiesa A, Santini P, Garlatti E, Luis F and Carretta S 2024 {\em Reports on
  Progress in Physics\/} {\bf 87} 034501

\bibitem{Toul77}
Toulouse G 1987 Theory of the frustration effect in spin glasses: {I} {\em Spin
  Glass Theory and Beyond\/} ed M{\'e}zard M, Parisi G and Virasoro M~A
  (Singapore: World Scientific) pp 99--103 reprinted form: G.~Toulouse 1977
  \textit{Communcations de Physique} \textbf{2} 115--9

\bibitem{Diep04}
Diep H~T and Giacomini H 2004 Frustration---{E}xactly solved frustrated models
  {\em Frustrated Spin Systems\/} ed Diep H~T (Singapore: World Scientific)
  chap~1, pp 1--58

\bibitem{Shas81}
{Sriram Shastry} B and Sutherland B 1981 {\em Physica\/} B+C {\bf 108} 1069--70

\bibitem{Voig95}
Voigt A, Richter J and Kr{\"u}ger S~E 1995 {\em Journal of Low Temperature
  Physics\/} {\bf 99} 381--3

\bibitem{Rich96}
Richter J, Voigt A, Kr{\"u}ger S~E and Gros C 1996 {\em \JPA\/} {\bf 29}
  825--36

\bibitem{Kami15}
Kamieniarz G, Florek W and Antkowiak M 2015 {\em Phys. Rev.\/} B {\bf 92}
  140411(R)

\bibitem{Flor16}
Florek W, Antkowiak M and Kamieniarz G 2016 {\em Phys. Rev.\/} B {\bf 94}
  224421

\bibitem{Flor19}
Florek W, Kamieniarz G and Marlewski A 2019 {\em Phys. Rev.\/} B {\bf 100}
  054434

\bibitem{Schm22}
Schmidt H~J and Schr{\"o}der C 2022 {\em Z. Naturforsch. A\/} {\bf 77} 1099 --
  1120

\bibitem{GraphTh}
Koh K~M, Dong F, Ng K~L and Tay E~G 2015 {\em Graph Theory\/} (Singapore: World
  Scientific)

\bibitem{Schm03}
Schmidt H~J and Luban M 2003 {\em \JPA\/} {\bf 36} 6351--78

\bibitem{Schm17a}
Schmidt H~J 2017 Theory of ground states for classical {H}eisenberg spin
  systems {I} arXiv:1701.02489

\bibitem{Hara53}
Harary F 1953 {\em The Michigan Mathematical Journal\/} {\bf 2} 143--146

\bibitem{ThMag}
Mattis D~C 2006 {\em The Thory of Magnetism Made Simple\/} (Singapore: World
  Scientific)

\bibitem{Giam11}
Giampaolo S~M, Gualdi G, Monras A and Illuminati F 2011 {\em Phys. Rev.
  Lett.\/} {\bf 107} 260602

\bibitem{Baez2018}
Baez M~L 2018 {\em Numerical methods for frustrated magnetism\/} Dissertation
  Freien Universit{\"a}t Berlin sec.\ 5.2.2

\bibitem{MatAlg}
Harville D~A 1997 {\em Matrix Algebra From a Statistician's Perspective\/} (New
  York, NY: Springer New York)

\bibitem{AnalIFC}
Fritzsche K and Grauert H 2002 {\em From Holomorphic Functions to Complex
  Manifolds\/} ({\em Graduate Texts in Mathematics\/} vol 213) (New York, NY:
  Springer New York)

\end{thebibliography}
\end{document}